\documentclass[preprint,superscriptaddress,preprintnumbers,prd]{revtex4}
\usepackage{latexsym}
\usepackage{amsthm}
\usepackage{amsmath}
\usepackage{graphicx}
\usepackage{amssymb}

\newcommand{\cm}{\ensuremath {\mathrm{~cm}}}
\newcommand{\s}{\ensuremath {\mathrm s}}

\def\OO{{\cal O}}

\newcommand{\TeV}{\,\mathrm{TeV}}
\newcommand{\GeV}{\,\mathrm{GeV}}
\newcommand{\MeV}{\,\mathrm{MeV}}
\newcommand{\keV}{\,\mathrm{keV}}

\newcommand{\fb}{\,\mathrm{fb}}

\newcommand{\be}{\begin{equation}}
\newcommand{\ee}{\end{equation}}
\newcommand{\bea}{\begin{eqnarray}}
\newcommand{\eea}{\end{eqnarray}}
\newcommand{\f}[2]{\ensuremath{\frac{#1}{#2}}}
\newcommand{\bef}{\begin{figure}[htbp]\begin{center}}
\newcommand{\eef}{\end{center}\end{figure}}

\newcommand{\parf}[2]{\left(\frac{#1}{#2}\right)}

\newcommand{\gsim}{\lower.7ex\hbox{$\;\stackrel{\textstyle>}{\sim}\;$}}
\newcommand{\lsim}{\lower.7ex\hbox{$\;\stackrel{\textstyle<}{\sim}\;$}}

\begin{document}
\pagestyle{plain}
\preprint{SLAC-PUB-14025}
\preprint{SU-ITP-10/14}
\title{\Large On the Origin of Light Dark Matter Species}
\author{Rouven Essig}
\email{rouven@stanford.edu}
\affiliation{Theory Group, SLAC National Accelerator Laboratory, Menlo Park, CA 94025}
\author{Jared Kaplan}
\email{jaredk@slac.stanford.edu}
\affiliation{Theory Group, SLAC National Accelerator Laboratory, Menlo Park, CA 94025}
\author{Philip Schuster}
\email{schuster@slac.stanford.edu}
\affiliation{Theory Group, SLAC National Accelerator Laboratory, Menlo Park, CA 94025}
\author{Natalia Toro}
\email{ntoro@stanford.edu}
\affiliation{Theory Group, Stanford University, Stanford, CA 94305}
\date{\today}
\begin{abstract}
TeV-mass dark matter charged under a new GeV-scale gauge force can
explain electronic cosmic-ray anomalies.  We propose that the CoGeNT
and DAMA direct detection experiments are observing scattering of
light stable states --- ``GeV-Matter'' --- that are charged under this
force and constitute a small fraction of the dark matter halo.  Dark
higgsinos in a supersymmetric dark sector are natural
candidates for GeV-Matter that scatter off protons with a universal
cross-section of $5\times 10^{-38} \cm^2$ and can naturally be split
by 10--30 keV so that their dominant interaction with protons is
down-scattering.  As an example, down-scattering of an $\OO(5)$ GeV
dark higgsino can simultaneously explain the spectra observed by both
CoGeNT and DAMA.  The event rates in these experiments correspond to a
GeV-Matter abundance of $0.2\!-\!1\%$ of the halo mass
density. This abundance can arise directly from thermal freeze-out at
weak coupling, or from the late decay of an unstable TeV-scale WIMP.
Our proposal can be tested by searches for exotics in the
BaBar and Belle datasets.
\end{abstract}
\maketitle
\tableofcontents
\section{Introduction}
For decades the WIMP --- a particle with TeV-scale mass that
freezes out through weak-strength couplings ---
has been the canonical dark matter candidate.  
The electronic cosmic ray excesses observed by 
PAMELA~\cite{Adriani:2008zr},
Fermi~\cite{Abdo:2009zk}, and
HESS~\cite{Collaboration:2008aaa,Aharonian:2009ah}, along with the
discovery of the WMAP and Fermi
``haze''\cite{Finkbeiner:2003im,Finkbeiner:2004us,Hooper:2007kb,Dobler:2007wv,Dobler:2009xz},
have both bolstered and complicated the WIMP picture.  Together these
experiments suggest that WIMP dark matter either annihilates or decays
into lepton-rich final states (see e.g.~\cite{Cirelli:2008pk}), with a rate 10-1000 times larger than
expected from $s$-wave annihilation.  Dark matter charged under a new GeV-scale
$U(1)_D$ `dark force' that kinetically mixes with hypercharge can explain these results through either Sommerfeld-enhanced annihilation 
\cite{Finkbeiner:2007kk, ArkaniHamed:2008qn, Pospelov:2008jd,
  Cholis:2008qq, Cholis:2008wq} or decay \cite{Ruderman:2009tj} into light dark-sector
states, which are in turn forced by kinematics to decay into leptons.  Such a GeV-scale sector is naturally realized in
models with supersymmetry-breaking near the weak scale, where
$D$-term mixing dynamically generates a GeV-scale mass for the $U(1)_D$
gauge boson $A'$  \cite{Dienes:1996zr,Cheung:2009qd,Katz:2009qq,Morrissey:2009ur,Cui:2009xq} (summarized in Section \ref{sec:model}).

The spontaneous breaking of this $U(1)_D$ gauge symmetry requires a higgs sector, and this sector can naturally contain stable GeV-scale states such as dark higgsinos.   
These and other stable ``GeV-Matter'' particles charged under $U(1)_D$ interact with protons with a cross-section 
\be
\label{DDCrossSection}
\sigma_{\chi, p} \approx 5\times 10^{-38} {\rm~cm}^2
\ee
that is {\bf completely fixed by Standard Model couplings}~\cite{Cheung:2009qd}.   Because this cross section is so large, even a small abundance of GeV-Matter can be observed at direct detection experiments.  In fact, we should expect GeV-Matter states to comprise only a small fraction of the dark matter halo because their annihilation cross-section is large ($\sim \alpha^2/m_{\mathrm{GeV}}^2$). 

\begin{figure*}[!]
\begin{center}     
\includegraphics[width=.485\textwidth]{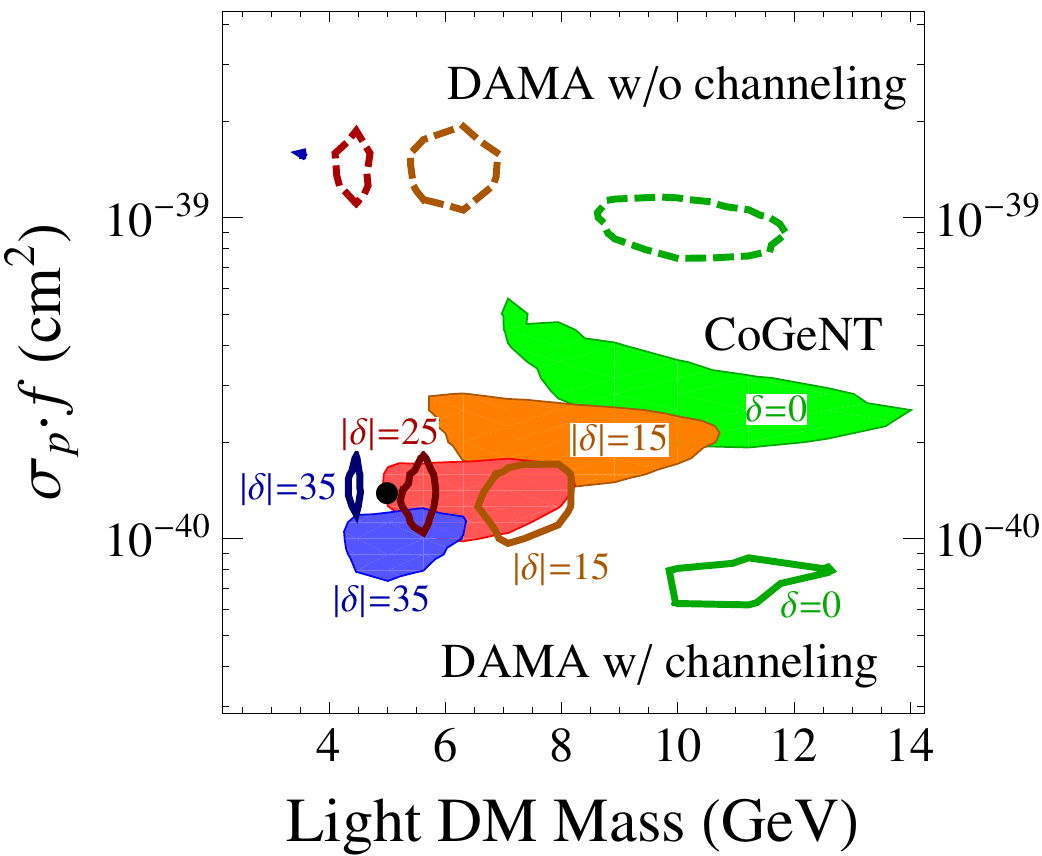}
\;
\includegraphics[width=.485\textwidth]{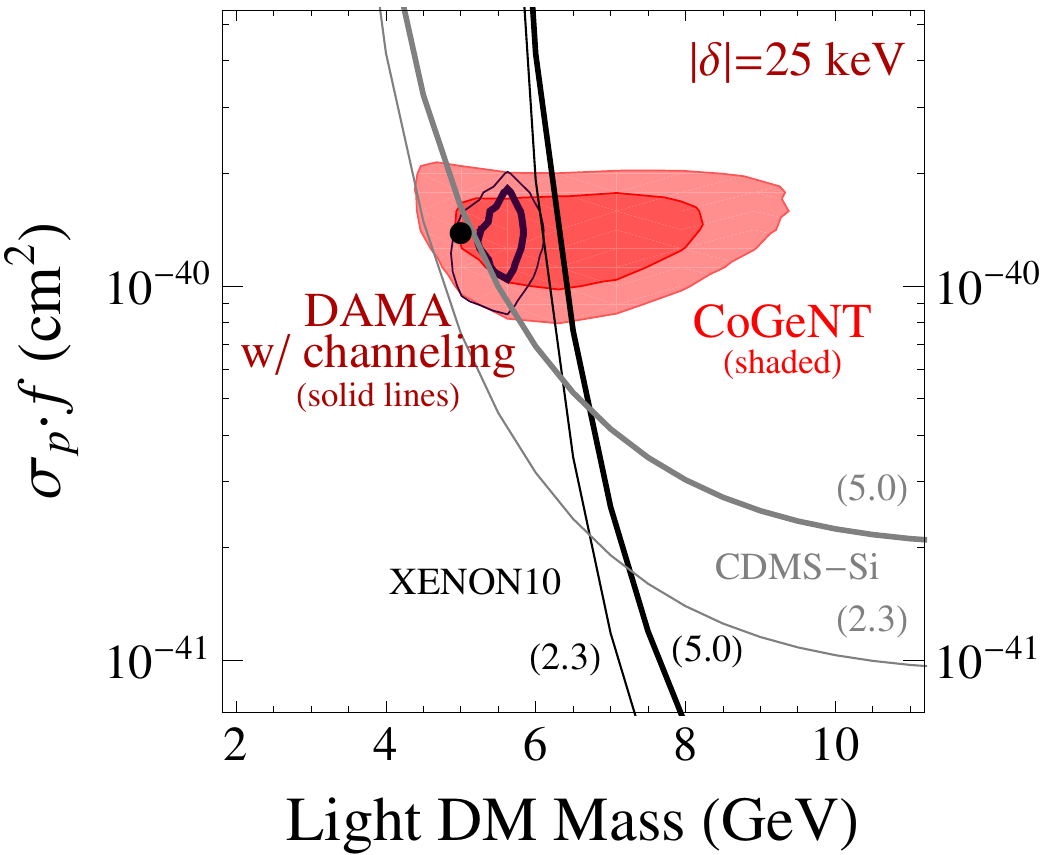}
\caption{ {\bf Left:} Regions of scattering cross-section per nucleon
  times GeV-Matter mass fraction in the halo versus mass favored by CoGeNT
  (filled regions), DAMA with channeling (solid-line contours), and
  DAMA without channeling (dashed contours) for light dark matter
  scattering coherently off protons.  We show favored regions (from
  right to left) for elastic scattering (green) and inelastic
  splittings $|\delta|=15$ keV (orange), $25$ keV (red), and $35$ keV
  (blue), with equal fractions of up- and down-scattering.  The
  contours correspond to ``$1.6\sigma$,'' neglecting systematic
  uncertainties.  
  The black dot and upside-down triangle correspond to benchmark 
  points we consider in Figure \ref{fig:DAMACogentfit}.
  {\bf Right:} A close-up of the $|\delta|=25$ keV
  allowed region with ``$3\sigma$'' contours added and constraints
  from XENON10 (black lines) and CDMS-Si (gray lines).  Thick (thin) lines denote 5.0 (2.3) events expected,
  again neglecting systematics.  More details are given in
 \S\ref{sec:DirectDetection}.}
\label{fig:DD1}
\end{center}
\end{figure*}

Two experiments have reported anomalous events that may arise from scattering of a light dark matter species. DAMA \cite{Bernabei:2000qi,Bernabei:2005hj,Bernabei:2008yi,Bernabei:2010mq} has reported a $9 \sigma$ annual modulation signal at 2--5 keVee in their NaI crystal detector.  
Early this year, the CoGeNT collaboration reported about 100 events from an unknown source in a low-threshold, high-resolution Ge detector \cite{Aalseth:2010vx}; these events are consistent with light dark matter scattering with a mass and cross-section similar to those needed to explain the DAMA modulation.  As shown in Figure \ref{fig:DD1} (see Section~\ref{sec:DirectDetection} for details), both signals can be  explained by 3--12 GeV dark matter scattering with a cross-section $\sigma_{\rm eff} \sim (1-5)\times 10^{-40}$ cm$^2$, or by a fraction of the halo scattering with a larger cross-section (see also \cite{Aalseth:2010vx,Fitzpatrick:2010em,Andreas:2010dz}).  If the events reported by these experiments arise from light dark matter scattering, they provide further evidence for a new sector at the GeV scale.  

In light of the cross section (\ref{DDCrossSection}), dark sector higgsinos at 3-12 GeV naturally explain the CoGeNT and DAMA excesses if they comprise a fraction
\be f_{\rm light} \equiv
\frac{\rho_{\rm light}}{\rho_{\rm DM}} \approx (2 {\rm - } 10)\times
10^{-3}
\label{flight}
\ee
of the halo mass density.   We propose two natural origins for an abundance of this size:
\begin{description}
\item[Thermal freeze-out] generates an abundance (assuming
  $s$-wave annihilation)
\be
\label{eq:annscaling}
f_{\rm light} \sim 3 \times 10^{-3} \parf{m_{\rm light}}{6 \GeV}^2
\parf{1/150}{\alpha}^2,
\ee
which yields the desired GeV-Matter relic density for small
couplings, as we discuss in Section \ref{sec:annihilation1}.
\item[Late Decays] of TeV-mass WIMP-sector states generate a density 
\be
\frac{n_{\rm light}}{n_{\rm heavy}} \sim 1, \qquad f_{\rm light} \sim
\frac{m_{\rm light}}{m_{\rm heavy}} \frac{\rho_{\rm heavy}}{\rho_{\rm
    DM}} \sim \rm{few} \times 10^{-3}
\ee
of GeV-Matter that can easily dominate over its thermal abundance.
This sharp prediction is insensitive to the details of the decay process, and agrees with the halo fraction required to explain the CoGeNT and DAMA signals!  The decaying
TeV-mass particle can be the scalar superpartner of a stable fermionic
WIMP, as discussed in Section \ref{sec:decay}.  
\end{description}
We therefore propose a new unified framework for interpreting both the direct detection and cosmic ray anomalies from the same dynamics.

$U(1)_D$-breaking mass splittings of ${\cal O}(\mbox{few
  keV}-\mbox{few GeV})$ for both WIMPs and GeV-Matter arise
from generic interactions with TeV-scale states.  In the
presence of such mass splittings, dark matter scatters
through inelastic up- and down-scattering, with elastic
processes highly suppressed.  These splittings have two important
effects: when the GeV-Matter states are split by ${\cal O}(10\mbox{
  keV})$, direct detection is dominated by down-scattering, which
improves the agreement between DAMA and CoGeNT and permits lower
masses of the light dark matter than in the elastic case, as discussed in
Section \ref{sec:DirectDetection} and more generally in
\cite{Graham:2010ca}.  A larger $\gtrsim$ MeV splitting of the
TeV-scale WIMP allows it to explain the
INTEGRAL 511 keV excess and renders it
practically invisible to direct detection experiments \cite{Strong:2005zx, Finkbeiner:2007kk}.

The success of these models is particularly striking in light of the difficulty of
generating such high direct detection rates through Standard Model
processes, even when a large abundance of light dark matter is generated through an asymmetry \cite{Kaplan:2009ag}.  Dark matter coupled to the $Z$ boson
can explain the reported rate, but 
is quite constrained by LEP measurements of the $Z$
invisible width~\cite{Kuflik:2010ah}.  Standard Model Higgs exchange, in contrast,
leads to typical cross-sections $\lesssim 10^{-43} \cm^2$ (see e.g.~\cite{Giedt:2009mr}) because the
Higgs couples only weakly to nuclei.  Explanations of the CoGeNT
signal through scalar interactions posit a new light scalar that mixes
significantly with the Higgs, but remains relatively light through a
percent-level fine-tuning \cite{Fitzpatrick:2010em,Andreas:2010dz}.
In contrast, the low mass of our dark matter and mediator arise from
simple dynamics.

\subsection*{Implications for $B$-factories}

Analysis of existing low energy collider data is crucially important for testing 
the GeV-Matter scenarios in this paper 
(see e.g.~\cite{Essig:2009nc,Batell:2009yf,Bossi:2009uw,AmelinoCamelia:2010me,Freytsis:2009ct,McDonald:2010iq}). 
The strongest constraint on $U(1)_D$ gauge boson decays to Standard Model
matter is from a BaBar search for a resonance in $\gamma \mu^+\mu^-$
in $\Upsilon(3s)$ data \cite{Aubert:2009cp}, which is sensitive to the
``radiative return'' process $e^+e^- \rightarrow \gamma A'$.  
The parameter space consistent with the GeV-Matter explanation of the
CoGeNT and DAMA excesses is just below the limit from this
search, but {\bf can be tested by a search in the full Belle and
  BaBar datasets}, which together comprise about a factor of 30 higher
statistics than the $\Upsilon(3s)$ data.  
The ``higgs$'$-strahlung'' process \cite{Batell:2009yf} may also be accessible at $B$-factories,
leading to a striking six-lepton final state.  
Searches at $J/\psi$- and $\phi$-factories can also explore the
low-mass parameter space \cite{Yin:2009mc,Li:2009wz,Bossi:2009uw,AmelinoCamelia:2010me}.

\subsection*{Down-Scattering and Lighter Dark Matter}
Three flaws have been noted with elastically scattering light dark
matter as an explanation of the DAMA annual modulation signal:
the predicted modulation spectrum grows exponentially at low energies while the DAMA spectrum appears to fall near threshold
\cite{Chang:2008xa}.  Moreover, the mass
range favored by both the DAMA and CoGeNT spectra, $\sim \! 10$ GeV, is
disfavored by  XENON10 \cite{Angle:2009xb, Manzur:2009hp, Angle:2007uj} and 
especially CDMS Silicon \cite{CDMS:talk} data (see
\cite{Fitzpatrick:2010em}).  Finally, the scattering rates required to explain the
CoGeNT and DAMA signals differ by about a factor of 5
(though this final discrepancy depends on the precise channeling properties 
of Iodine \cite{Fitzpatrick:2010em,Drobyshevski:2007zj, Bernabei:2007hw, MarchRussell:2008dy}).

However, there is no reason to expect the light dark matter scattering to be purely elastic.  Mass splittings of order 10-30 keV for GeV-Matter are readily generated by TeV-suppressed higher-dimension operators.  

Because the dark sector kinetically decouples from the Standard Model
at temperatures high compared to this splitting, both states will be
equally populated and the heavy state will be cosmologically long-lived
\cite{Finkbeiner:2009mi, Alves:2009nf, Alves:2010dd}.  The excited states interact with nuclei through down-scattering, which pushes the nuclear recoil spectrum to higher energies for a given dark matter mass.  As we discuss in
\S\ref{sec:DirectDetection}, this effect 
improves the agreement of light dark matter with DAMA modulation data (relative to the elastic case), 
brings the CoGeNT and DAMA expected cross-sections together, and is
consistent with the XENON10 null search.  The residual tension of this
scenario with the 5-tower CDMS Silicon analysis~\cite{CDMS:talk} is
extremely sensitive to the experimental selection efficiency near
threshold and small energy-scale systematics.  There is also tension with the DAMA unmodulated spectrum in \cite{Bernabei:2008yi}.  A general analysis of
down-scattering light dark matter \cite{Graham:2010ca} also finds a
lower-mass region consistent with DAMA and CDMS Silicon results.

\section{An Approximately Supersymmetric Dark Sector}\label{sec:model}
Our starting point is a 
simple extension
(\cite{Morrissey:2009ur,Cui:2009xq,Cheung:2009qd,Katz:2009qq}) of the
supersymmetric $SU(3)\times SU(2)_L\times U(1)_Y$ Standard Model that
dynamically generates a mass scale of ${\cal O}(\GeV)$ for the ``dark
force.''  The dark sector consists of two chiral superfields $h_u$ and
$h_d$ that are oppositely charged ($\pm 1$) under a new $U(1)_D$ gauge
group, but neutral under the Standard Model gauge groups.  No Standard
Model fields are directly charged under $U(1)_D$.  We also add a gauge
singlet $S$, and consider the following superpotential for the
GeV-scale sector:
\be
{\cal W}  = \lambda S h_u h_d - \frac{1}{4} W_D^2 - \frac{\epsilon}{2} W^{\alpha}_Y W_{\alpha D}.
\label{eqn:lightW}
\ee
Here, $W_D$ and $W_Y$ are the $U(1)_D$ and hypercharge vector
superfields, and the last term is the supersymmetric version of  
the usual gauge kinetic mixing,
\be\label{eq:kinmix}
\mathcal{L} = \f{\epsilon}{2} \, F_{Y,\mu\nu} F_D^{\mu\nu}. 
\ee 
Besides inducing the gauge kinetic mixing (\ref{eq:kinmix}) and 
gaugino kinetic mixing, the last term also induces a fixed $\epsilon$-suppressed
potential coupling of Standard Model and dark-sector scalar fields
through the mixing of Standard Model and dark-sector $D$-terms.
After electroweak symmetry breaking, this mixing generates a
dark-sector scalar potential
\be
V = \lambda^2 \left( |S h_u|^2 + |S h_d|^2 + |h_u h_d|^2 \right) + \frac{g_D^2}{2} (|h_u|^2 - |h_d|^2 + \zeta^2)^2,
\ee
that includes a negative mass-squared for the scalar $h_d$ (or $h_u$
if $\zeta^2 <0$), where
\be
\zeta^2 = \frac{\epsilon g_Y |\cos{2\beta}|}{2 g_Dg_2^2}m_W^2 \label{eq:zeta}
\ee
is the hypercharge $D$-term vev.

The stable vacuum is one where $U(1)_D$ is broken but supersymmetry is
approximately preserved (supersymmetry-breaking corrections, discussed below, are higher-order in
$\epsilon$).  The light fields are grouped into massive gauge and chiral supermultiplets, with masses
\be 
m_{A'}=m_{\tilde \lambda} = \sqrt{2} g_D \zeta \quad \mbox{and} 
\quad m_h = m_{\tilde h} = \lambda \zeta. 
\label{eq:dterm}
\ee 
The uneaten real part of $h_d$ and $\tilde h_d$ (which gives a Dirac mass to the gaugino $\tilde \lambda$) are in the massive gauge supermultiplet, while the $h_u$ and $S$ scalars and their superpartners $\tilde h_u$ and $\tilde S$ marry into the massive chiral supermultiplet.
The $A'$ mediates scattering of particles with $U(1)_D$-charge $q_d$  off Standard Model matter \textbf{with a non-relativistic scattering cross-section almost completely determined by Standard Model parameters},
\be\label{eq:DtermScatter}
\sigma_{\chi,p} 
 = 16\pi q_d^2 \frac{ \mu_{\chi,p}^2 \alpha_D \alpha c_W^2\epsilon^2}{m_{A'}^4}
 = \frac{(16\pi q_d^2 \alpha_2^2c_W^4)\mu_{\chi,p}^2}{(\cos{2\beta})^2m_W^4}
\approx 5 \times 10^{-38}  \left( \frac{2 q_d}{\cos{2\beta}} \right)^2\mbox{cm}^2,
\ee
with all dependence on $\epsilon$ and the dark gauge coupling $g_D$
canceling and only mild dependence on the ratio $\tan\beta$ of
Standard Model Higgs vevs and on the dark matter mass (we 
chose $m_\chi = 5$ GeV in the numerical estimate above).  

In addition to these light fields, we assume throughout that the
dominant WIMP component of dark matter arises from a vector-like pair of chiral
superfields $\Phi_{u,d}$ with charge $\pm q$, so that the full renormalizable superpotential is
\be
{\cal W_{\rm tot}}  = \lambda S h_u h_d - \frac{1}{4} W_D^2 - \frac{\epsilon}{2} W^{\alpha}_Y W_{\alpha D} 
+ M \Phi_u \Phi_d,
\label{eqn:heavyW}
\ee
with $M$ at the TeV-scale (for example, $M$ and the $\mu$-term of the MSSM Higgs fields may have the same origin).
The stability of $\Phi$ and absence of other couplings to the light fields can easily be guaranteed by global symmetries, in particular including a ${\mathbb Z}_2$ under which the $\Phi_{u,d}$ fields are odd.  Alternately, the stability of $\Phi$ is automatic if its charge $q$ is non-integer. 

\subsection*{Supersymmetry-Breaking Effects}
Supersymmetry-breaking effects are crucial for dark sector physics; we will always take fermionic states as 
our dark matter candidates, so we will insist that they are stable.  
In the light GeV-Matter sector, some scalar 
density is innocuous but it is important that the $\Phi$ scalars decay.

We begin by discussing the supersymmetry-breaking effects that arise inevitably from gauge mediation through the kinetic mixing and from anomaly mediation, and then we will consider further effects that can arise through gravity or direct mediation to the dark sector.

In models of gauge mediation, all scalars charged under $U(1)_D$ receive soft masses mediated from the messengers through kinetic mixing of order \cite{Zurek:2008qg,Morrissey:2009ur}
\be 
{\tilde m}^2_{GMSB}  \sim \epsilon^2 \frac{\alpha_D}{\alpha_Y} m_{E^c}^2, 
\ee 
where
$m_{E^c}$ is the mass of the right-handed selectron in the Minimal Supersymmetric Standard Model 
(MSSM).   A \emph{negative}
supersymmetry breaking mass-squared for $S$ is generated by RG evolution \cite{Morrissey:2009ur}, 
although it is suppressed by an additional loop factor times a potentially large logarithm.  The $R$-breaking 
$A h_u h_d S$ and $B \Phi_u \Phi_d$ terms are 
suppressed by ${\cal O}(\alpha_D\epsilon^2)$ relative to the MSSM bino mass, $m_{\tilde B}$. 
A Majorana soft mass for the dark gaugino is given by $\epsilon^2 m_{\tilde B}$, while the 
MSSM bino and dark gaugino have a mixing mass of $\epsilon m_{\tilde B}$.  In the absence of $U(1)_D$ symmetry-breaking, these would preserve a zero mass eigenvalue.  When $U(1)_D$ is broken they only induce a tiny splitting $\epsilon^2 m_{\tilde\lambda}^2/m_{\tilde B}$ of the Dirac gauginos into two Majorana states 
(approximately $(\tilde \lambda \pm \tilde h_d)/\sqrt{2}$), one heavier and
the other lighter than the $A'$.  
Both the lighter and heavier of these states can decay through kinetic mixing 
with the MSSM bino to $\gamma \tilde G$, with a lifetime~\cite{Morrissey:2009ur} 
\be
\tau_{\tilde \lambda} \sim 10^3 ~\s 
\left( \frac{\sqrt{F}}{10^{5}\GeV} \right)^4 
\left( \frac{1 \GeV}{m_{\tilde \lambda}} \right)^5
\left( \frac{10^{-3}}{\epsilon} \right)^2
\label{eqn:gauginoDecay}
\ee
as long as $m_{\tilde \lambda} > m_{3/2}$.  Constraints from BBN \cite{Jedamzik:2006xz} and from
EGRET searches for $\gamma$-ray lines from the galactic center require \cite{Pullen:2006sy}
either very low $\sqrt{F} \lesssim 10^{5}$ GeV for prompt gaugino decays or
rather high $\sqrt{F} \sim 10^{10}$ GeV, so that the gravitino is heavier
than the gaugino and the decay is forbidden.  In the latter case, the dark sector soft masses 
inevitably receive anomaly-mediated contributions \cite{Randall:1998uk,Giudice:1998xp} of order 
\be
\tilde m_{\rm AMSB} \sim \frac{\alpha_D}{4\pi} m_{3/2} \sim 1-100~{\rm MeV},
\ee
which are comparable to the gauge-mediated contributions to the scalar soft masses and dominate the gaugino soft mass.  

If $F$ is low, dark matter annihilation or decay to $A'$s at a rate
compatible with the FERMI $e^+e^-$ excess could potentially
generate a gamma-ray signal visible with existing or upcoming satellite data.
This is because supersymmetric couplings will also allow dark matter
annihilation/decay to $\tilde \lambda$, whose subsequent decay produces a
photon. This signal is fairly model-dependent --- it is reduced if 
dark matter annihilates or decays through supersymmetry breaking couplings, or if 
the gaugino decays primarily through $R$-parity-violating interactions.
Nonetheless, it would be interesting to investigate these signals in future work. 

Although the effects we considered above are irreducible, it is plausible to include additional supersymmetry breaking operators such as
\be
\int d^4 \theta \frac{X^\dag X}{M_*^2} S^\dag S   \ \ \ \mathrm{and} \ \ \ \int d^2 \theta \frac{X}{M_*}  W_D^2,
\ee
where $X$ is a supersymmetry-breaking-spurion with an $F_X$ component vev.  We can think of these terms 
as originating from `gravity mediation' with $M_* \sim M_{pl}$, but it is equally natural for $M_*$ to be much 
smaller.  These operators can contribute to the Majorana gaugino mass and also raise the scalar higgs sector 
($S$ and $h_u$) soft masses.  
In particular, the scalar $S$ can obtain a net positive supersymmetry-breaking-mass, forbidding the 
decay of the lightest higgsino to $S$ and a gravitino. 

\subsection*{Stable States and Their Properties}

We will always take our GeV-Matter (the light stable states) to be fermionic, so in this section we will be interested in the properties of these fermions and in the mechanisms through which we avoid stable scalars.

The Dirac gaugino \emph{can} be stabilized on cosmological timescales
if the supersymmetry-breaking $F$ is large, so that either the lifetime ($\propto F^2$) of the gaugino 
is very large or $m_{3/2}>m_{\tilde \lambda}$ (which requires $\sqrt{F} \gtrsim 0.5 \times 10^{10}$ GeV) 
so that it is kinemtically forbidden to decay into the gravitino.  
Nonetheless, it is not a good candidate 
to explain the CoGeNT and DAMA signals in our scenario.  
For such high $F$, the 
gaugino anomaly-mediated soft mass $\sim 10-100$ MeV is sufficiently large that
the heavier gaugino 
decays rapidly to $e^+e^-$ and the lighter gaugino.  Though the lighter component 
can comprise a non-negligible fraction of the dark matter, it can only
scatter elastically (because the excited state is depopulated and
up-scattering is kinematically forbidden) with a cross-section that is velocity-suppressed and 
thus too small to explain the direct detection signals.  

The higgs supermultiplet thus contains the only viable GeV-Matter
candidates within the model \eqref{eqn:lightW} that can explain CoGeNT
and DAMA.  We assume that the higgs supermultiplet is stabilized on
cosmological timescales by a discrete or continuous global symmetry.
We are primarily interested in higgsino GeV-Matter ($(\tilde h_u \pm \tilde S)/\sqrt{2}$), which readily
acquires a small inelastic splitting as we discuss below.  
The higgsinos are stable under either of two conditions: either $S$ must receive a positive soft-mass 
contribution that makes it heavier than the higgsino, or $F$ must be sufficiently large that decays  
$\tilde h\rightarrow S + \tilde G$ are kinematically forbidden.

If $F$ is low and the supersymmetry breaking contribution to the $S$ scalar mass is
positive (which requires some contribution besides gauge mediation),
all scalars can decay to their fermionic partners and a gravitino with a lifetime given 
parametrically by $\tau \sim 16\pi F^2/( m_h^4 \Delta m)$, and
likewise gauginos can decay to $\gamma \tilde G$ according to
\eqref{eqn:gauginoDecay}.  The higgsino is then the \emph{only} stable
relic in the light dark sector.  In this case, the scalar $\Phi$ also typically decays to $\tilde \Phi + \tilde G$.  

Large $F$ also leads to a stable higgsino if the gravitino mass is large enough to kinematically forbid 
higgsino decays to $S+\tilde G$ or $S+ \tilde h_d$.  In this scenario, one or both scalars in the higgs multiplet 
may be stable, but they are less phenomenologically important than the higgsinos.  
In particular, $S$ is neutral under $U(1)_D$ and therefore only has a very small direct detection 
cross-section induced by its mass mixing with $h_u$.  
Even at high $F$, a heavier $h_u$ scalar can typically decay through the same mass mixing to $S$ 
and an off-shell 
$A'$.  Even if the $h_u$ is cosmologically stable, it typically has no inelastic splitting, making it less 
observable in direct detection than the higgsinos which, as we will discuss below, typically down-scatter.

 A final possibility that we will not explore further in this work is that the higgsinos are actually lighter than the gaugino and gravitino states, so that the higgsinos are stabilized by R-parity.

Although supersymmetry-breaking effects leave the higgsino and $\tilde
\Phi$ as exactly degenerate Dirac fermions (in contrast with the gaugino),
interactions with 1-10 TeV states can lead to splittings by ${\cal
  O}(10\mbox{ keV}-1\mbox{ MeV})$ when these heavy states are
integrated out, for example, through the higher-dimension
superpotential operator
\be
\frac{1}{M_*} h_d^2 h_u^2.
\label{eqn:splitop}
\ee
Elastic scattering processes will be velocity suppressed and therefore negligible in all cases, so the higgsinos will only be visible to direct detection experiments through inelastic down-scattering.

In many models (see Appendix \ref{app:splittings}), the same
interactions generate an ${\cal O}(\MeV)$ or larger splitting for the
heavy dark matter fermions $\tilde \Phi_{u,d}$ and a smaller ${\cal O}(10 \mbox{
  keV})$ splitting for the higgsinos; the simplest example is to take
$q=1/2$ and include the marginal superpotential operator $h_u
\Phi_d^2$.  For splittings $\delta M\gtrsim 1$ MeV, the excited state of the
TeV-scale dark matter decays to the lower-mass state and an $e^+e^-$
pair, depopulating the initial state of the down-scattering process,
and up-scattering off nuclei is kinematically forbidden.
In this case, the leading direct detection cross-section would be elastic,
but it is suppressed by both $\left ( \frac{\delta M}{M} \right )^2 \sim 10^{-12}$ - $10^{-6}$
and by a factor of the non-relativistic velocity squared $\sim 10^{-6}$, making it completely negligible. 
For splittings $\sim 1$ MeV the $\tilde \Phi_{u,d}$ self-interactions may explain 
the INTEGRAL 511 keV excess\cite{Finkbeiner:2007kk}.

We note that it is less generic in these sectors to generate inelastic
splittings for the scalar $\Phi$ than for the $\tilde \Phi$ fermion, so
we insist that the scalar $\Phi$ decay.  At low $F$ the gauge-mediated
splittings are sufficient to allow decays $\Phi \rightarrow \tilde
\Phi+ \tilde G$.  For the higher-$F$ scenario and heavier gravitinos,
it is plausible that the $\Phi$ couples more directly to SUSY-breaking
than do the fields in the low-mass dark sector (for example, if its
$\mu$-term is related to SUSY-breaking), so that decays to the
gravitino are still kinematically allowed.  Alternately, in Section
\ref{sec:decay} we consider interactions with an intermediate-scale
supersymmetry-breaking sector that lead to $\Phi$ scalar decays but
keep the fermion $\tilde \Phi$ cosmologically stable.

\section{GeV-Matter from Thermal Freeze-Out}
\label{sec:higgsSectorDM}
Given the proximity of the CoGeNT and DAMA mass scales to the
GeV-scale, a natural possibility is that the higgsinos of the dark sector are themselves stable and give
rise to observable direct detection signals.  We focus
on $\lambda \gtrsim \sqrt{2} g_D$, so that the higgs supermultiplet is
heavier than the gauge supermultiplet, and we will assume that global symmetries guarantee that this multiplet is stable on time-scales of order the age of the universe.  Note that the gaugino states could be cosmologically stable, but they are not a
viable candidate to explain the CoGeNT and DAMA signals as we discussed in Section \ref{sec:model}.  
The higgsino relic abundance agrees with \eqref{flight} in a small-coupling
limit. For larger couplings the thermal higgsino abundance is small but, as we will see in the next section, the late decay of a TeV-mass WIMP gives rise to a relic density of higgsinos of the appropriate size.

We will also discuss how an additional light vector-like species (not part of the dark gauge or higgs sector) is a viable GeV-matter candidate if it is charged under $U(1)_D$. 

\subsection*{Dark Higgsino Freeze-Out}
Higgsino annihilation during freeze-out is dominated by annihilation to the real scalar $h_d$ and a longitudinal $A'$, with cross-section
\be
\sigma_{\tilde h} = \frac{5 \pi \alpha_\lambda^2}{8 m_{\tilde h}^2 v}
\ee
where $\alpha_\lambda = \lambda^2/(4\pi)$.
This dominates over all other annihilation modes, which are either $p$-wave suppressed or suppressed by $g^2/\lambda^2$, leading to an abundance
\be
f_{\tilde h} \equiv \frac{\Omega_{\tilde h}}{\Omega_{DM}} \simeq 3.0 \times 10^{-3} \left(\frac{1/150}{\alpha_\lambda} \right)^2 \left(\frac{m_{\tilde h}}{6 \GeV} \right)^2,
\ee
where $\alpha_\lambda = \lambda^2/(4\pi)$, and we have chosen the baseline $m_{\tilde h} = 6$ GeV 
near the CoGeNT and DAMA best-fit regions for $25 \keV$ inelastic splitting (see 
Section \ref{sec:DirectDetection}).  
Using \eqref{eq:DtermScatter} with $q_d=1/2$, this gives rise to an effective direct-detection cross-section 
\bea\label{eq:themoneyHiggsino}
\sigma_{\rm eff} \equiv f_{\tilde h} \sigma_{\tilde h, p} &\simeq& 1.5\times 10^{-40} \mbox{cm}^2 
\\ \nonumber
&& \times \left( \frac{1}{\cos{2\beta}} \right)^2 
 \left(\frac{1/150}{\alpha_\lambda} \right)^2 \left(\frac{m_{\tilde h}}{6 \GeV} \right)^2,
\eea
Normalizing this to the $\sigma_{\rm eff}$ derived from DAMA and CoGeNT data determines  the dark-sector higgs vev within this setup (at around 20 GeV), and the gauge coupling as a function of the $A'$ mass:
\be
\alpha_D = \f{1}{2} \alpha_\lambda \parf{m_{A'}}{m_{\tilde h}}^2 \simeq 10^{-4} \parf{m_{A'}}{1\GeV}^2  \parf{m_{\tilde h}}{6 \GeV}^{-1} \parf{\sigma_{\rm eff}}{1.5 \times 10^{-40} \cm^2}^{-1/4}
\ee
near the CoGeNT/DAMA best fit.  A separation of scales between the $A'$
and higgsino masses, as suggested by the lepton-rich cosmic ray
signals, implies a rather small dark gauge coupling in this scenario.
This in turn can be translated into a prediction for $\epsilon$ using
\eqref{eq:zeta} and \eqref{eq:dterm},
\be
(\epsilon c_W)^2 \simeq 1.9 \times 10^{-5} \times \parf{m_{A'}}{1 \GeV}^2 \parf{m_{\tilde h}}{6 \GeV}^{1/2} \parf{\sigma_{\rm eff}}{1.5 \times 10^{-40} \cm^2}^{1/4}.
\label{ecw:annihilate}
\ee
If the $A'$ decays directly to $e^+e^-$, models that produce
$\sigma_{\rm eff}$ as a thermal relic are very near the sensitivity limit
of existing $B$-factory searches discussed in Section
\ref{sec:collider}.  

For larger gauge couplings, the relic abundance of the higgsinos from thermal freeze-out is 
too small to explain the CoGeNT and DAMA signals.  
However, as we will discuss in Section \ref{sec:decay}, a relic abundance of the desired size can easily be generated by other means.

\subsection*{Vector-Like GeV-Matter}\label{sec:annihilation1}

Now let us consider a slightly less minimal model that exhibits the parametrics of equation (\ref{eq:annscaling}) without a direct link between the GeV-Matter mass and the $A'$ mass.  Consider a light vector-like species $\chi_\pm$ with superpotential 
$W \supset \mu_\chi \chi_+ \chi_-$ and $\mu_\chi \sim 5$-$10$ GeV.  Like $\Phi_{u,d}$, the $\chi_\pm$ states may be stabilized by a discrete symmetry such as a $Z_2$ where $\chi_\pm \to - \chi_\pm$.  
A higher-dimension coupling of $\chi$ to the higgs-sector field $h_{d}$ such as 
\be
\frac{1}{\Lambda} \chi_+^2 h_d^2
\ee
will lead to a $\sim 10$ keV splitting for the fermionic light dark
matter states $\tilde \chi_\pm$ when $\langle h_d \rangle \sim 0.1\!-\!1$
GeV and $\Lambda \gtrsim 1\!-\!10$ TeV.  Note
that the scalar states in the $\chi$ multiplets can decay to $\tilde
\chi + \tilde G$ or $\tilde \chi + \tilde \lambda$ pairs depending on the
gravitino and gauge supermultiplet masses.

In a minimal scenario, the $\chi$ relic abundance is set by the annihilation cross-section to lighter higgs and gauge-multiplet states, which we parametrize as
\be
\sigma_{\tilde \chi} \simeq \frac{3 \pi \alpha_D^2}{2 m_\chi^2 v}
\ee
This gives rise to a relic density
\be
f_{\tilde \chi} \equiv \frac{\Omega_{\tilde \chi}}{\Omega_{DM}} \simeq 6\times 10^{-4} \left(\frac{1/100}{\alpha_D} \right)^2 \left(\frac{m_{\chi}}{6 \GeV} \right)^2.
\ee

The  $\chi$ scatter in direct detection experiments with total cross-section \eqref{eq:DtermScatter} with $q_d=1$, giving rise to an effective cross-section 
\bea\label{eq:themoneyVectorlike}
\sigma_{\rm eff} \equiv f_\chi \frac{\Omega_{\tilde \chi}}{\Omega_{DM}}\sigma_{\tilde h, p} &\simeq& 1.2\times 10^{-40} \mbox{cm}^2 
\\ \nonumber
&& \times \left( \frac{1}{\cos{2\beta}} \right)^2
\left(\frac{1/100}{\alpha_D} \right)^2 \left(\frac{m_{\chi}}{6 \GeV} \right)^2.
\eea
The same parametrics applies to the heavy $\Phi_{u,d}$ (again, assuming that annihilation through gauge interactions dominates), albeit with a
higher $g_*$ in the freeze-out calculation.  For example, $\alpha_D = 1/100$ used above would lead to $\Omega_\Phi = \Omega_{DM}$ for $\mu_\Phi = 550$ GeV. The above model is similar to that discussed in \cite{Cholis:2009va}, but we consider a different hierarchy of masses and couplings. 

\section{GeV-Matter from the Decay of Heavy WIMPs}
\label{sec:decay}

In this section, we will present a simple mechanism that can simultaneously explain the cosmic ray data and the direct detection signals.  Our mechanism populates the GeV-Matter states via the late decay of a TeV-scale WIMP-sector particle, resulting in a relic abundance that precisely fits the CoGeNT and DAMA direct detection rates (with the scattering cross-section in Eq.~(\ref{eq:DtermScatter})); it explains the cosmic-ray data through the decay of a TeV-scale WIMP with a lifetime on the order of $10^{26}$ s.  Both of these effects originate from the same supersymmetric operator.  

We consider a setup where the fermionic WIMP and its scalar superpartner freeze out with comparable relic densities. 
While the WIMP itself will have a lifetime greater than the age of the Universe, its scalar partner will decay more rapidly into the dark higgsino state, providing the dominant contribution to the GeV-Matter density, as we now show.  If we denote the mass of the light and heavy states by $m_L$ 
and $m_H$, respectively, then the relic density of the light and heavy states from freeze-out are 
$\rho_{L {\rm (f.o.)}}\!\! \propto \sigma_{L {\rm (ann.)}} \propto \! m_L^2$ and $\rho_{H {\rm (f.o.)}}\!\!  \propto \sigma_{H {\rm (ann.)}} \propto \! m_H^2$, respectively, so that $\rho_{L {\rm (f.o.)}}/\rho_{H {\rm (f.o.)}} \sim m_L^2/m_H^2$ 
for roughly equal couplings.  
After one of the heavy states decays into the light state, the light state will inherit 
the \emph{number} density of the heavy state, 
$n_L\! \sim \! n_H = \rho_{H {\rm (f.o.)}}/m_H\! \propto \! m_H$, so that the relative relic density from decay $\rho_L/\rho_H \sim m_L/m_H \gg m_L^2/m_H^2$ 
dominates over $\rho_{L {\rm (f.o.)}}$.  Moreover, for $m_L\! \sim \! 5$ GeV and $m_H\! \sim \!1\!-\!2.5$ TeV, the stable heavy state can dominate the 
dark matter relic density so 
$\rho_L/\rho_{\rm DM} \! \sim \! m_L/m_H\! \sim \! (2\!-\!10) \times 10^{-3}$, which is in precise accord with the  
DAMA and CoGeNT rates (see Eq.~(\ref{flight})) if the light state has a direct detection cross-section 
given by (\ref{eq:DtermScatter}). 

For a concrete example of this scenario, we again consider the low-energy superpotential
\be\label{eq:lowenergy}
{\cal W} = -\f{1}{4} W_D^2  - \frac{\epsilon}{2} W^{\alpha}_Y W_{\alpha D} + 
\lambda S h_u h_d + M \Phi_u \Phi_d  + \f{1}{2 M_1} h_d^2 h_u^2 + \f{1}{2 M_2} h_d^2 \Phi_u^2
\ee
where the $S$, $h_u$, and $h_d$ are GeV-scale states and the fermions $\tilde \Phi_{u,d}$ are the TeV-scale WIMPs.  The two higher-dimensional operators 
induce the necessary splittings among the WIMPs and GeV-Matter, so that in direct 
detection experiments the GeV-Matter can down-scatter and the heavy dark matter 
component is forbidden from scattering.  Now let us include the K\"ahler potential terms 
\be\label{eq:Kahler}
\int d^4 \theta \frac{1}{M^2_*} \Big(\Phi_u^\dag e^V h_u X^\dag Y + \Phi_u^\dag e^V h_u X Y^\dag + {\rm h.c.}\Big),
\ee
where $X$ and $Y$ are heavy fields in the supersymmetry breaking sector, and $V$ is the 
$U(1)_D$ vector field.  When $X$ and $Y$ acquire non-zero $F_X$ and $F_Y$ -terms, we break two $\mathbb{Z}_2$ symmetries 
under which both superfields $\Phi_u$ and $\Phi_d$ and either $X$ or $Y$ are charged, allowing the 
$\Phi_u$ scalar to decay through mass mixing with the $h_u$ state. 
Any non-zero $B_\mu$-term, generated, for example, from gauge mediation, will maximally mix the 
$\Phi_u$ and $\Phi_d$ states, and also allow the $\Phi_d$ scalar to decay.  
Two residual (non-supersymmetric) accidental $\mathbb{Z}_2$'s under which the \emph{fermions} $\tilde\Phi_u$ and 
$\tilde\Phi_d$ and either the $X$ or the $Y$ \emph{scalars} are charged, are only broken by the scalar vevs  
$\langle X\rangle$ and $\langle Y\rangle$, and induces $\tilde\Phi_{u,d}$ fermion decay through kinetic mixing with $h_u$.  The fermions are parametrically longer-lived than the scalars, because the higher mass dimension of $F_X,\,F_Y$ must be compensated for by the TeV-scale $\Phi$ mass.

The dominant $\Phi$ scalar decay channels are then 
$\Phi \rightarrow h_u+A'$, $\Phi\rightarrow \tilde h_u+\tilde \lambda$, and especially
$\Phi\rightarrow \tilde h_d+\tilde S$ due to the fact that $\lambda > g_D$; we estimate the
lifetime as
\bea
\tau_{\Phi} &\sim& \left(\frac{4\pi}{\alpha_\lambda} \right ) \frac{M^4_*M^3}{|F_XF_Y|^2}
\\ \nonumber
&\sim& 10^4 \ \mbox{s} \ \left( \frac{1/30}{\alpha_\lambda}\right ) \left ( \frac{M_*}{10^{19}\GeV} \right )^4 \left ( \frac{3\times10^7 \GeV}{\sqrt{F}} \right )^8 \left( \frac{M}{2 \TeV} \right)^3,
\eea
where we have set $F_X=F_Y=F$.
The dominant $\tilde \Phi$ fermion decays are
$\tilde \Phi\rightarrow \tilde h_u+A'$, $\tilde\Phi \rightarrow h_u+\tilde \lambda$, $\tilde \Phi\rightarrow \tilde h_d^{\dagger}+S^*$,
and $\tilde \Phi\rightarrow h_d^*+\tilde S^{\dagger}$
with lifetime
\bea
\tau_{\tilde \Phi} &\sim& \left(\frac{1/30}{\alpha_\lambda} \right ) \frac{M^4_*}{|\langle X \rangle \langle Y \rangle |^2M}
\\ \nonumber
&\sim&10^{26} \ \mbox{s} \ \left( \frac{1/30}{\alpha_\lambda}\right ) \left ( \frac{M_*}{10^{19}\GeV} \right )^4 \left ( \frac{(2\times10^6 \GeV)^2}{\langle X \rangle \langle Y \rangle } \right )^2 \left( \frac{2 \TeV}{M} \right).
\eea
It is important to this argument that less suppressed operators
involving $X$ and/or $Y$ and K\"ahler terms involving chiral products
such as $\Phi_u h_d$ are forbidden (as in the UV completion in
Appendix \ref{sec:uvcompletion}) \emph{or} that one of the scalar vevs
$\langle X \rangle,\,\langle Y \rangle$ vanishes.  In either of these
situations, the interactions in \eqref{eq:Kahler} are the leading
decays. 
In general, all that is required to populate the higgsino states is for 
scalar $\Phi$ decay to occur, so fermion $\tilde \Phi$ decays can be rendered
negligible if $\langle X\rangle \ll \sqrt{F}$ or $\langle Y\rangle \ll \sqrt{F}$. 

The scalar $\Phi$ decays occur well after the dark higgs and gauge sectors freeze out, so 
that the abundance of the GeV-Matter and potentially long-lived 
(depending on the gravitino mass) $\tilde \lambda, \tilde h_d$ are determined by the density of $\Phi$.
Though initially energetic, the $\Phi$ decay products redshift and become non-relativistic 
before matter-radiation equality.  
Consequently, the velocity profile of these decay products in 
galaxy halos is controlled by gravitational infall like that of the WIMP dark matter.  
A lifetime in the range $10~\s \lesssim \tau_\Phi \lesssim 10^5~\s$  conservatively avoids BBN constraints for $\Omega_\Phi h^2 \simeq 0.1$ and assuming that $\Phi$ decays never produce baryons; if the $\Phi$ branching fraction to Standard Model leptons is small ($\OO(10^{-2})$ is readily achieved from ratios of couplings) then shorter and longer lifetimes are allowed \cite{Jedamzik:2006xz}.  

The fermion decays $\tilde \Phi\rightarrow \tilde h_u+A'$, $\tilde \Phi\rightarrow h_u+\tilde\lambda$ and $\tilde \Phi\rightarrow h_d^*+\tilde S^{\dagger}$
produce leptons through the decay $A'\rightarrow \ell^+\ell^-$, $h_d \rightarrow \ell^+\ell^-$,  
and $h_u\rightarrow S+A'\to S+\ell^+\ell^-$ or $h_u\rightarrow S+\ell^+\ell^-$ (the latter decay is through 
an off-shell $A'$ if $m_{h_u}\!-\!m_{s} < m_{A'}$). 
Consequently, we expect that this model can explain the PAMELA 
and FERMI signals as arising from $\tilde \Phi$ decay for 
$\langle X\rangle \! \sim \! \langle Y\rangle \! \sim \! \sqrt{F_X}\! \sim \! \sqrt{F_Y} \! \sim \! 10^{3}\!-\!10^4 \TeV$, 
with a possible higher-energy component of the cosmic ray spectrum arising from Sommerfeld-enhanced 
annihilations (which by itself may not suffice to explain the signal \cite{Feng:2009hw}). 
Note that the $\tilde\Phi$ decays also produce high-energy gauginos, which can decay into high energy 
gravitinos and photons, $\tilde \lambda\rightarrow \gamma+\tilde G$.  If this decay is fast, 
then it may generate a photon signals close to current limits 
(see e.g.~\cite{Meade:2009mu,Essig:2009jx,Abdo:2010nc}).  
This constraint is trivially avoided if the gravitino is heavy and/or $\tilde \lambda$ is long-lived. 
(Note that if the gaugino is unstable, its lifetime should  also be $\lesssim 10^5$ s to avoid constraints from photodissociation of nuclei \cite{Kawasaki:2004qu,Jedamzik:2006xz})
That dark matter decay induced by higher-dimensional operators can produce sizable cosmic ray signals
was first pointed out in \cite{Eichler:1989br}, and more recently considered in \cite{Nardi:2008ix,Arvanitaki:2008hq}.  

We present a UV completion of the above model \eqref{eq:lowenergy} and \eqref{eq:Kahler} in 
Appendix \ref{sec:uvcompletion}.

\section{Direct Detection Phenomenology}\label{sec:DirectDetection}

In this section, we discuss direct detection signals from 
light dark matter species at DAMA and CoGeNT, and the
constraints from XENON-10 and CDMS-Si.  Our focus is on extending previous analyses \cite{Aalseth:2010vx,Fitzpatrick:2010em} to models of light dark matter with an inelastic splitting
$\delta\lesssim$ 35 keV and where down-scattering naturally dominates,
which 
(i) changes the recoil and modulation spectra, especially at low recoil energies,
(ii) allows lower dark matter masses (down to 3--4 GeV),
(iii) permits excellent simultaneous fits to DAMA and CoGeNT spectra and rates, and
(iv) greatly reduces the strength of limits from XENON10, and magnifies the sensitivity of CDMS-Si limits to systematic uncertainties.
Down-scattering at lower dark matter masses has also been proposed as
an explanation of the DAMA signal \cite{Bernabei:2008mv,Graham:2010ca}, and
\cite{Graham:2010ca} analyzes the phenomenology of a variety of
down-scattering signals in detail.

\subsection{Kinematics of Inelastic Down-Scattering}
\label{ssec:downscat}

Vector-current interactions of fermions with a small mass
splitting are always inelastic.  When one component of a Dirac fermion
$\Psi=(\psi_L\, \bar \psi_R)$ of mass $m$ receives a small Majorana
mass $\delta$, it splits $\Psi$ into two 
components $\psi_{1,2}$ with Majorana masses $\approx m \pm \delta/2$,
which are approximately $\psi_1 = (\psi_L + \psi_R)/\sqrt{2}$ and
$\psi_2 = i (\psi_L - \psi_R)/\sqrt{2}$.  Upon rotation of the Dirac
fermion's gauge interactions to the mass eigenstate basis, all
couplings $ g_{ij} A^\mu \bar \psi_i \bar \sigma_\mu \psi_j $ are
generated.  If the initial state's coupling is vector-like, the
coefficients $g_{ii}$ are suppressed by $\delta/m$ (as expected
because these couplings break the $U(1)_D$ gauge symmetry, and
$\delta$ parametrizes the $U(1)_D$ breaking felt by the fermion), but
if $\psi_L$ and $\psi_R$ have different chiral couplings, the $g_{ii}$
are generically non-zero.  An additional suppression of elastic scattering exists \emph{irrespective of the size 
of $g_{ii}$}, since the $A'$ couplings to matter are very nearly pure
vector-like (with axial couplings down by $(m_{A'}/m_Z)^2$ and
proportional to the nuclear spin rather than its total charge), and
since the matrix element for a Majorana fermion scattering off a vector
current vanishes at zero velocity.  Thus, even for the chiral case,
\emph{elastic scattering cross-sections are always suppressed by
  $v^2$}.

Every halo particle in the high-mass state can down-scatter (unlike up-scattering, there is no threshold kinetic energy).  However, the recoil kinematics of
down-scattering differs markedly from that of elastic scattering.  In
the following, we briefly describe the kinematics of 
down-scattering events for a dark matter particle of mass $M_\chi$
scattering off a nucleus of mass $m_N$.  If the dark matter is
incident at velocity $v$, the center-of-mass frame has a velocity
$v_{\rm cm} = \mu v /m_N $ relative to the lab frame, where $\mu$ is the 
dark matter-nucleus reduced mass.  The squared
velocity of the recoiling nucleus in the center-of-mass frame,
assuming the dark matter mass increases by $\delta$ in scattering
(i.e. $\delta <0$ for down-scattering) is simply
\be
v_{out,{\rm cm}}^2 = \frac{\mu^2}{m_N^2} \left(v^2 - \frac{2\delta}{\mu}\right).
\ee
Therefore, the average lab-frame recoil energy is
\be
\bar E(v) = \frac{1}{2} m_N \langle v_{lab}^2 \rangle 
= \frac{1}{2} m_N  \left(v_{\rm cm}^2 + v_{out,{\rm cm}}^2\right)
= \frac{\mu^2}{m_N} \Big(v^2 + \frac{\delta}{\mu}\Big),
\ee
while the range of recoil energies is given by
\be
\frac{\mu^2}{2 m_N}  \left( v - \sqrt{v^2 - \frac{2 \delta}{\mu}} \right)^2 \ \leq \ E_R \ \leq \ \frac{\mu^2}{2 m_N} \left(v + \sqrt{v^2 - \frac{2 \delta}{\mu}} \right)^2
\ee
Thus, when $|\delta| \gtrsim \mu v^2$, the effect of down-scattering on
the nuclear recoil spectrum is dramatic, with mean nuclear recoil energy
\be
\langle E_R  \rangle \approx \frac{|\delta| \mu}{m_N}
\label{Erbar}
\ee
and width $\Delta E_R \sim (\mu/m_N) \sqrt{8 \delta \mu v^2}$.  

Because the recoil spectrum is not peaked at zero, experiments with
low energy thresholds relative to $|\delta|\mu/m_N$ are sensitive to a
large fraction of scattering events, even if the scattering WIMP is
quite light (an elastically scattering state of roughly the same mass,
such as a dark higgs, can have a comparable halo abundance but go
un-noticed because its detection efficiency is typically much lower).
On the other hand, experiments with high energy thresholds, which are
only sensitive to the tail of the recoil energy distributions, are
less sensitive to down-scattering than to elastic scattering with the
same characteristic scale for $E_R$.  This is because, in the elastic
case, the typical $E_R$ and its exponential tail are both determined
by $(\mu^2/m_N) v^2$.  Down-scattering models achieve the same
characteristic $E_R$ scale with lower dark matter masses, and
therefore a sharper exponential fall-off in rate at high recoil
energies.  Therefore, the importance of differences in energy
thresholds between experiments is amplified.  Moreover, the bounds
from these experiments are extremely sensitive to uncertainties in the
threshold energy and the detection efficiency near threshold.

An approximate figure of merit for comparison of thresholds across
different target nuclei (when $m_{DM} \ll m_N$) is the threshold
\emph{recoil momentum} 
\be 
p_* \equiv \sqrt{2 E_{R,thresh}  m_N}.
\ee 
Reduced mass factors further decrease the sensitivity of
detectors using the lightest nuclei, but for $m\sim 5$ GeV the effect
is small, even for Silicon. The nominal CoGeNT, CDMS-Si, and XENON10
thresholds correspond to $p_*=16.0$, $19.2$, and $32.2$ MeV
respectively.  Thus, the increased threshold-dependence of
downscattering dark matter will significantly weaken the XENON10
constraint on downscattering models.  The impact on nominal limits
from CMDS-Si is quite mild, as its recoil momentum threshold is within
$30\%$ of CoGeNT's.  However, the CDMS-Si efficiency near threshold
\cite{CDMS:talk} is small and varies rapidly.  The above comparison
makes clear that any uncertainty in near-threshold acceptance affects
CDMS-Si sensitivity dramatically.

\subsection{Direct Detection Scattering Rate}
The differential rate for spin-independent scattering of a dark matter
species $\chi$ off a target containing $N_T$ nuclei of mass $m_N$ is given by 
\bea
\f{dR}{dE_R} & = & N_T \f{m_N}{2 \mu_n^2} \f{f_{\chi} \rho_{\rm dm}}{m_\chi} 
\f{(f_p Z + f_n (A-Z))^2}{f_p^2}  \sigma_{\chi,p} \nonumber \\
& & 
\times F^2(E_R)  \int _{v_{\rm min}}^{v_{\rm esc}} d^3 v \f{f(\vec{v}-\vec{v}_e)}{v} ,
\label{eqn:dRdEr}
\eea
where $E_R$ is the energy of the recoiling nucleus, $\mu_n$ is the dark matter-nucleon (proton or neutron) reduced mass, 
$\rho_{\rm dm}  \simeq 0.3 \GeV/\cm^3$ is the local dark matter
density, $m_\chi$ and $f_{\chi}$ are the mass and fractional abundance
$\rho_\chi/\rho_{dm}$ of $\chi$, 
$Z$ and $A$ are the atomic number and atomic mass of the target 
nucleus, $f_p$ and $f_n$ are the effective coupling strengths of the dark matter particle to protons 
and neutrons, respectively (for kinetic mixing, $f_n=0$ so the
scattering rate is proportional to $Z^2$), and 
\be
\sigma_{\chi,p} \equiv \frac{\mu_n^2 }{64 \pi^2 m_{DM}^2 m_p^2} \int d\Omega_{cm} |{\cal M}|^2
\ee 
is the effective scattering cross-section of the dark matter particle
off protons (this equals the physical scattering cross-section in the
elastic limit, but differs for inelastic scattering by a ratio of
outgoing to incoming velocities in the CM frame).  The parametrization
of the form factor $F(E_R)$,  dark matter velocity distribution
$f(\vec v - \vec v_r)$, and $E_R$-dependent velocity threshold
$v_{min}$ in the last two terms of \eqref{eqn:dRdEr} are
described in Appendix \ref{app:DirectDetection}.

Throughout this section, we treat $\sigma_{\chi,p} f_\chi$ as the
parameter to be fit or constrained by event rates in direct detection
experiments.  For the inelastic scenarios, $f_\chi$ refers to the
\emph{total} $\chi$ mass fraction, which we assume to be equally
divided between the excited and ground states, so that the fraction of
the halo that can down-scatter is $f_\chi/2$, while the fraction that
can up-scatter is $f_\chi/2$ times a Boltzmann factor, and we add the
two contributions.

\subsection{The Data}
\paragraph*{CoGeNT}
CoGeNT is an ultra-low-noise germanium detector with a fiducial mass
of 330 g that operated for 56 days in the Soudan Underground
Laboratory with an ionization energy threshold of 0.4 keVee (keV
electron equivalent, i.e.~ionization energy).  Between 0.4 keVee and
3.2 keVee, the detector recorded an approximately uniform background
rate, sharp cosmogenic peaks, and about 100 events of unknown origin
between 0.4 and 1.0 keVee, peaked near threshold
\cite{Aalseth:2010vx}.  Background explanations of the latter
component are not excluded, but its spectrum is sufficiently similar
to that expected from light dark matter scattering off germanium to
warrant further study 
\cite{Aalseth:2010vx,Fitzpatrick:2010em,Feldman:2010ke,Kuflik:2010ah,Andreas:2010dz}.

The ionization energy $E_{\rm ee}$ (in keVee) is related to the nuclear recoil of energy $E_R$ by a 
quenching factor.
%
Following \cite{cogentNote}, we use a fit to the Lindhard model with $k=0.2$,
\be
\label{quenching}
E_{\rm ee} = 0.20 \left( \frac{E_{R}}{1 \ \mathrm{keV}} \right)^{1.12}.
\ee
Assuming this quenching factor, the threshold energy is 
$0.4~{\rm keV_{ee}} = 1.9~{\rm keV_r}$ and $p_* = 16.0$ MeV.

The proximity of the putative signal to the detector's noise threshold
and the similarity of its shape to that of a hypothetical exponential background pose a challenge for interpretation
of the CoGeNT dataset.  
We do not fit to the first bin (0.4-0.45
keVee) on the grounds that it is within $1\sigma$ energy resolution
from the noise threshold and may be significantly contaminated.
We fit three contributions to the remaining bins
between 0.45 and 3.2 keVee:
\begin{enumerate}
\item A uniform background with free normalization.
\item Gaussian peaks centered at 1.1 keVee and 1.29 keVee, originating
  from known radioactive decays, each of which is modeled as a
  Gaussian with free normalization and width given by the detector
  energy resolution $\delta E \simeq \sqrt{(0.0697)^2 + 0.000977\,
    E_R}$ keV from \cite{Aalseth:2008rx}.
\item A contribution from dark matter scattering calculated according
  to \eqref{eqn:dRdEr}, using the time-average earth velocity and the quenching factor \eqref{quenching}.  We do not include any possible
channeling in Ge.  We neglect the finite energy resolution as it is
much narrower than the intrinsic energy distribution of dark matter
recoils.
\end{enumerate}
All expectations are multiplied by the detection efficiency plotted in
\cite{Aalseth:2008rx} as a function of energy.  As the number of
events in each bin is small, we use Poisson log-likelihood rather than
$\chi^2$ to quantify goodness of fit.  It is important to remember
that the CoGeNT low-energy excess is entirely consistent with an
exponentially falling background.  However, we do not find it useful to include such a
component when fitting the spectrum because it is not
distinguishable from a dark matter signal with the present
experimental statistics.  At higher statistics, a spectral
feature (which is expected in much of the down-scattering parameter
space) could help to establish the dark matter hypothesis if it is
correct.  

\paragraph*{DAMA/NaI and DAMA/LIBRA}
The DAMA/NaI \cite{Bernabei:2005hj} and DAMA/LIBRA
\cite{Bernabei:2008yi,Bernabei:2010mq} experiments have reported an
annual modulation in the rate of single-hit events with ionization
energy in the interval 2--5 keVee, with a consistent phase over 13
years.  Here we fit this signal assuming it originates from dark
matter scattering off the Sodium and Iodine nuclei in their detectors.

The detector response to nuclear recoils is complicated by the
``channeling effect'' \cite{Drobyshevski:2007}, whereby a fraction of
recoiling nuclei deposit almost all of of their energy electronically.
Following the parametrization of \cite{Fairbairn:2008gz}, we take
$E_{ee} = E_R$ in a fraction $f_{\rm chan}$ of scattering events (the
``channeled events''), and \eqref{quenching} in the remainder, with
\be 
f_{\rm chan,Na} \simeq \f{e^{-E_R/18}}{1+0.75 E_R}, \quad f_{chan, I} \simeq 
\f{e^{-E_R/40}}{1+0.65 E_R}, \quad {\rm QF(Na)} = 0.3, \;\mbox{ and } {\rm QF(I)}
= 0.09 \label{chanfunc}.
\ee
Thus, some fraction of events at $2-5$ keVee arise from $2-5$ keV
nuclear recoils. In the models we consider, channeled I recoils
dominate the expected signal, with a smaller contribution from
un-channeled Na recoils.

It is important to note that the functional forms in \eqref{chanfunc}
are parametrizations of the Monte Carlo results in
\cite{Bernabei:2007hw}, which differ significantly from analytical
models \cite{Gelmini:2009he,BGGtoAppear} and have not been tested
experimentally in the relevant energy range.  Moreover, 
partial channeling is not included in these
parametrizations.  All of our quantitative predictions for the DAMA
experiment depend very sensitively on the behavior of channeling in the 2-5
keV range.

We calculate the modulation spectrum by subtracting the expected
energy spectrum when the Earth's velocity in the galactic frame is at
its annual minimum, from the expected spectrum when this velocity
is at its annual maximum.  We smear this spectrum with a gaussian whose
width corresponds to the DAMA energy resolution, $\delta E =
0.0091+0.448/\sqrt{E}$ \cite{Bernabei:2008yh}.

\paragraph*{XENON10}

The XENON10 experiment used a liquid Xenon target with an exposure of 316 kg-days, and saw no events consistent with light dark matter.  
We use the 2--20 keV nuclear recoil energy bin and the reported software 
cut acceptances and nuclear recoil band acceptance in \cite{Angle:2009xb} to set a limit.  

A large source of uncertainty comes from the scintillation efficiency,
$L_{\rm eff}$, which corresponds to the amount of nuclear recoil
energy that gets recorded as scintillation light.  This affects the
detector's threshold energy, and a small change in $L_{\rm eff}$ has a
large impact on the resulting constraints.  Unless otherwise stated,
we use the recent measurement of $L_{\rm eff}$ in
\cite{Manzur:2009hp}, which yields weaker constraints than
the value $L_{\rm eff}=0.19$ quoted in \cite{Angle:2009xb}.  Our fit to this $L_{\rm eff}$ curve implies a true energy threshold of $4.2~{\rm keV_r}$, and $p_* = 32.2$ MeV.

\paragraph*{CDMS-Si}
The CDMS experiment contains both Germanium and Silicon detectors.  Far less data has been collected for the Silicon detectors, but we have checked that this data nonetheless sets a tighter constraint than the recent Germanium analysis \cite{Ahmed:2009zw}.
For CDMS-Si, a total exposure of 53.5 kg-days has been obtained with 
a threshold of 7 keV nuclear recoil \cite{CDMS:talk} (we use this updated result 
rather than the results in \cite{Akerib:2005kh}), corresponding to $p_* = 19.2$ MeV.  
We use the efficiency provided in \cite{CDMS:talk}.  
While the resulting constraints generate tension with the CoGeNT and DAMA preferred 
regions, they are extremely sensitive to the precise threshold energy, as we will demonstrate.

\subsection{Results}
Figure \ref{fig:DD1} shows the regions consistent with the CoGeNT and
DAMA data for different choices of the mass splitting $\delta$ from 0
to 35 keV.  We take $f_p=1$ and $f_n=0$, as appropriate for dark
matter interacting with nuclei through a kinetically mixed mediator. 
The elastic scattering regions (green region and contours)
are consistent with the results
for $f_p=f_n=1$ in \cite{Aalseth:2010vx,Fitzpatrick:2010em}, after
rescaling our results by a factor of ($Z$(Ge)/$A$(Ge))$^2 \simeq
0.19$.  The CoGeNT regions correspond to regions where $-2\Delta\ln
\mathcal{L} + 2 \ln \mathcal{L}_{\max} = 6.25$ (and 14.16 for the
right plot), where $\mathcal{L}$ is the Poisson likelihood and
$\mathcal{L}$ is its maximum value for \emph{any} $|\delta|$ (here
near $\delta=0$).  Likewise the DAMA contours correspond to $\chi^2 -
\chi^2_{\rm min} = 6.25$ (14.16) relative to the overall minimum
$\chi^2$, occurring near $|\delta|=20$ keV.  These boundaries
approximate formal 90\% ($1.6\sigma$) and 99.7\% ($3\sigma$)
likelihood regions, but they neglect systematic uncertainties in
the dark matter halo and detector response that are known to be
important.

We note that the down-scattering parameter space allows simultaneous fits to the CoGeNT and DAMA data.  
In our parametrization of channeling, the preferred region is 
\be
m_{\rm dm} \sim 4-10~{\rm GeV}, \;\;  \sigma_0\times f  \sim (1-2) \times 10^{-40}~{\rm cm}^2, \;\; 
|\delta| \sim 15-35~{\rm keV}.
\ee
but would shift according to any change in the treatment of channeling.  

\begin{figure*}[!]
\begin{center}     
\includegraphics[width=.485\textwidth]{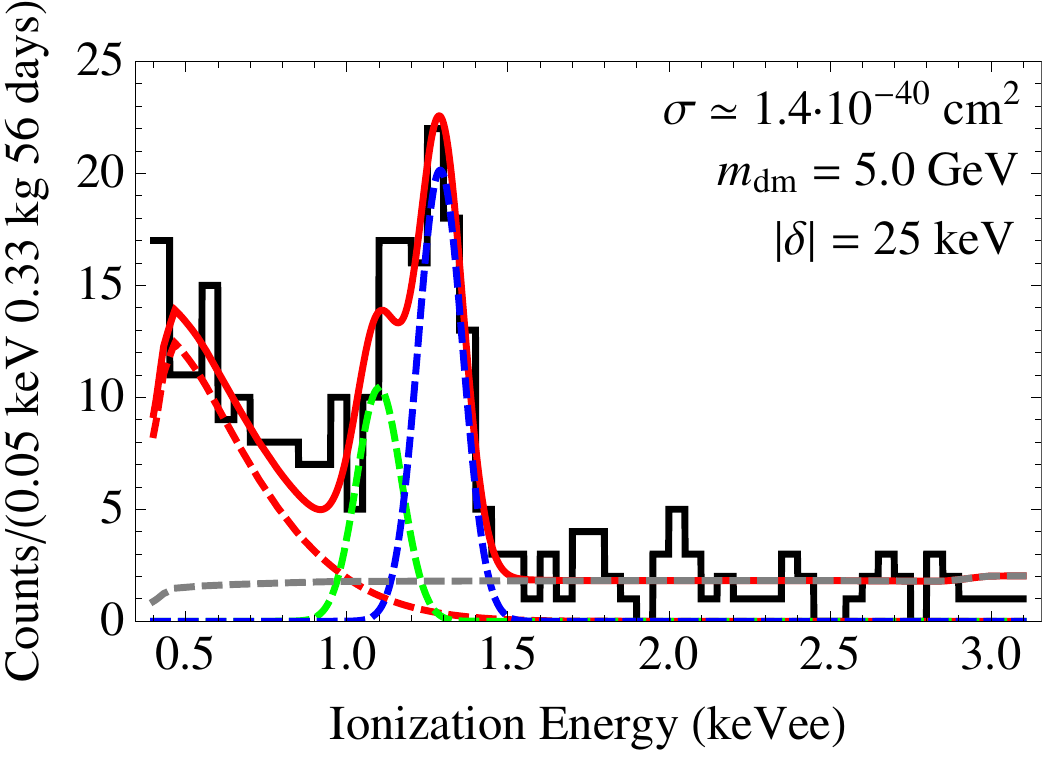}
\;\;
\includegraphics[width=.485\textwidth]{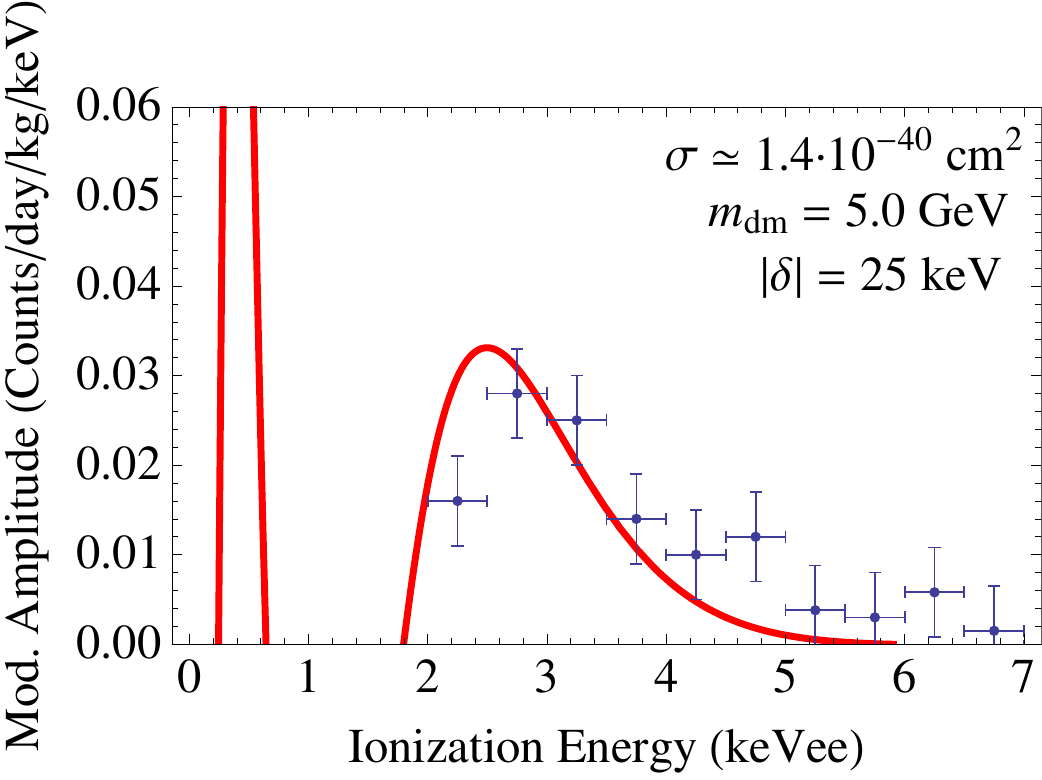}
\caption{
A fit (solid red line) to the CoGeNT data (left) and DAMA data (right).  In the CoGeNT fit, we assume an energy-dependent detection efficiency \cite{Aalseth:2010vx} and include four contributions: 
dark matter (red dashed), two gaussians for the cosmogenic peaks near 1.1 keV and 
1.29 keV (green and blue dashed), and a constant background (gray dashed).  
The dark matter mass (5.0 GeV), cross-section ($1.4\times 10^{-40}$ cm$^2$), and splitting 
$\delta=25$ keV correspond to a benchmark point that fits both data sets very well 
(indicated by the dot in Figure \ref{fig:DD1}).
The expected number of events in CDMS-Si and XENON10 is about 4.3 and 0, respectively.   
Assuming a 20\% larger energy threshold in CDMS-Si would lead to 1.1 events expected.  
}
\label{fig:DAMACogentfit}
\end{center}
\end{figure*}

Figure \ref{fig:DAMACogentfit} shows expected spectra at CoGeNT and
DAMA for a benchmark point, $|\delta| = 25$ keV, $m_{\rm dm}=5.0$ GeV, and $\sigma \cdot f 
\simeq 1.4\times 10^{-40}$ cm$^2$, a point in this preferred region
(indicated by the dot in Figure \ref{fig:DD1}).  The CoGeNT recoil
spectrum agrees with \eqref{Erbar} after accounting for the quenching
factor.  The modulation at DAMA comes from the
high-energy tail of the nuclear recoil energy distribution, and is
dominated by channeled iodine recoils (with a smaller contribution
from unchanneled sodium recoils).  The large component below the 2--5
keVee DAMA region of interest (and below the 2 keV threshold of the
DAMA modulation analysis) comes from unchanneled iodine scatters.  For
these parameters, we expect about 4 events in CDMS-Si and none in XENON10 under
our baseline assumptions.  Under different assumptions about the DAMA
detector response, the preferred regions move (and can reduce the
tension with CDMS-Si and XENON10).  

\paragraph*{Constraints}

\begin{figure*}[!]
\begin{center}     
\includegraphics[width=.485\textwidth]{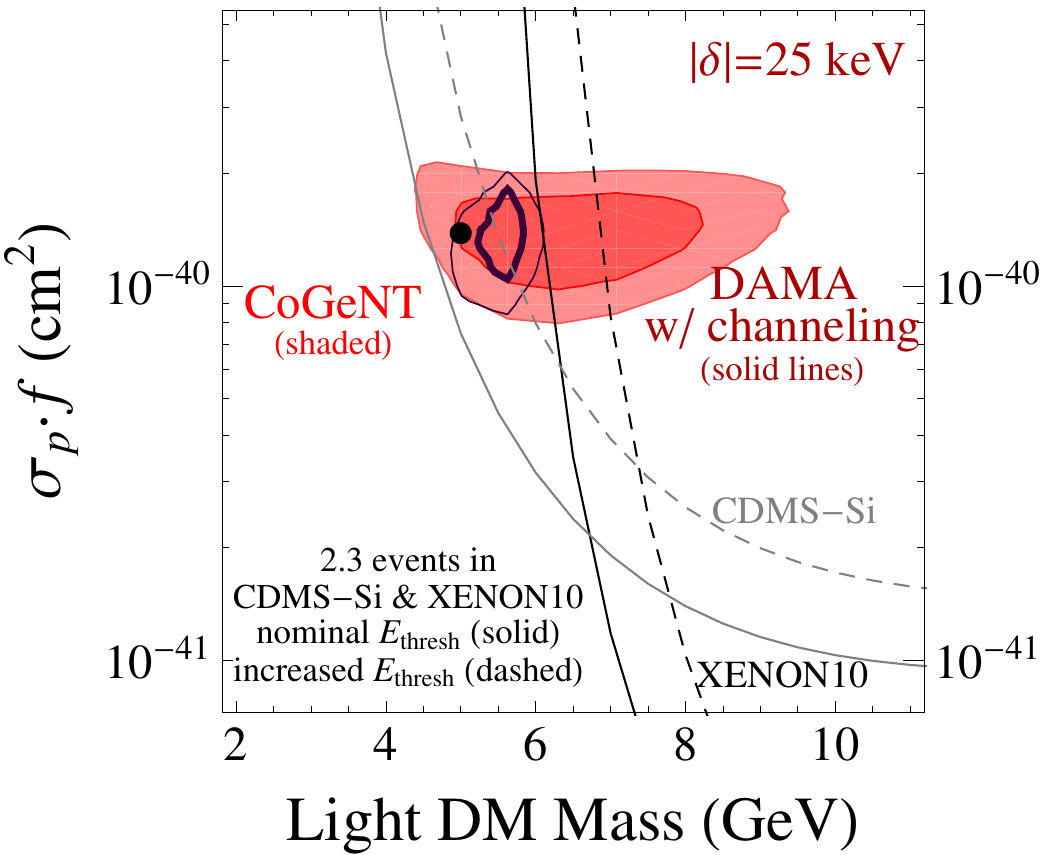}
\;\;
\includegraphics[width=.485\textwidth]{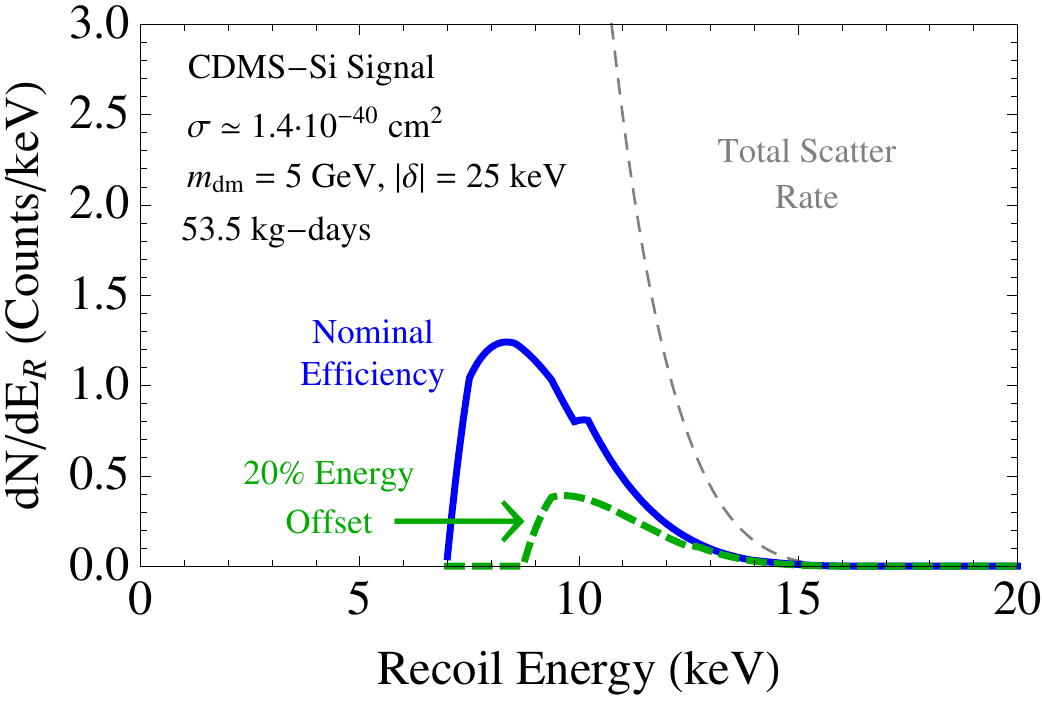}

\caption{The effect of energy thresholds on XENON10 and CDMS-Si
  Limits.  \textbf{Left:} The contours and thin solid lines are as in
  the left panel of Figure \ref{fig:DD1}.  The dashed black and gray
  lines indicate the change in XENON10 and CDMS-Si sensitivity when
  their threshold energy is increased by 20\%.  In the case of
  CDMS-Si, we use the same rescaling of recoil energies in determining
  the detection efficiency.  \textbf{Right:} The spectra in CDMS-Si
  for the benchmark point with $\delta=25 \keV$ (top panels of Figure
  \ref{fig:DAMACogentfit}), assuming the nominal efficiency (blue solid curve) or assuming a 20\%  offset of the energy threshold and energy-dependent efficiency (dashed green curve).  The total scatter rate with no efficiency correction applied (thin dashed gray curve) is also shown for reference.
}
\label{fig:thresh}
\end{center}
\end{figure*}

CDMS-Si, XENON10, and the DAMA unmodulated single-hit spectrum all
have the potential to constrain these models.  However, significant
systematic uncertainties apply to each of these limits.  Curves
corresponding to 2.3 and 5 expected events in XENON10 and CDMS-Si
data, with the baseline assumptions detailed above, are included in
the right panel of Figure \ref{fig:DD1}.  These limits, especially the
limit from CDMS-Si, exhibit tension with the preferred overlapping
region of DAMA and CoGeNT.  However, their severity is rather
sensitive to various assumptions about each detector's response near
threshold, while the precise regions preferred by CoGeNT and DAMA are
vulnerable to similar uncertainties as well as energy-scale
uncertainties (and particularly, in the case of DAMA, to the detailed
features of channeling).  We illustrate the sensitivity of the XENON10
and CDMS-Si limits to the energy scales of these experiments in Figure
\ref{fig:thresh}, where we have varied the true energy scale of their
detector thresholds by 20\%.

The potential importance of the DAMA unmodulated single-hit rate
\cite{Bernabei:2008yi} (which varies slowly between 1 and 1.5
counts/kg/day for energies above 1.5 keVee and rises rapidly at lower
energies) in constraining models of light dark matter was emphasized
in \cite{Chang:2008xa}.  With the standard channeling assumptions, our
benchmark points overpredict the total rate near 2 keVee by about a
factor of 2. This is a potentially severe source of tension, but it
is driven by two poorly constrained features of channeling --- the behavior of channeling 
at low ($\lesssim 3$ keVee) energies, and a possible energy dependence of the quenching factor
in this range. Blocking and de-channeling effects can dramatically reduce the channeling 
fraction at low energies, rendering the predicted single-hit rate consistent \cite{Gelmini:2009he}. 
A careful study of channeling in the few-keV
region is needed to make sharp predictions of these rates and spectral profiles
in order to settle this issue.
The single-hit tension is also driven by fitting the \emph{dip} in the lowest DAMA bin.  Allowing for
a mild uncorrected decrease in efficiency in this bin (so that the
rate in this bin is comparable to the 2nd bin) would render lower
splittings consistent with the data, and make the unmodulated spectrum
consistent.  
We also note that the lower-mass parameter regions discussed in \cite{Graham:2010ca}, which
must assume Ge channeling to explain the CoGeNT excess, do not have
this problem.  

Finally, although we have checked that the preferred regions and constraints depend only mildly on the mean halo velocity, and are quite insensitive to variations in $v_{\rm esc}$, other changes to the halo velocity profile may have 
a more significant impact.

\section{$B$-Factory Predictions}
\label{sec:collider}

For all of the scenarios outlined above, the 
mass of the vector $A'$ and higgs states are underneath the 
energy threshold for direct production at the $B$-factories BaBar
and Belle \cite{Essig:2009nc, Batell:2009yf}. 
The main production processes are $e^+e^-\rightarrow A' \gamma$
and $e^+e^-\rightarrow A' h_d$. 
For $m_{A'}\sim 1 \GeV$, the dominant 
decay modes of $A'$ are $e^+e^-$, $\mu^+\mu^-$ or $\pi^+\pi^-$, all of which can have branching fractions 
$\gsim 10\%$.
Other decays with less than $\lesssim 10\%$ branching fractions include,
$\pi^0\omega$, $K^+K^-$, $\pi^+\pi^-2\pi^0$, $2\pi^+2\pi^-$, $\pi^0\pi^+\pi^-$, and $K^0\bar{K^0}$.
The $h_d$ will mainly decay to an $A'$ and two soft leptons or pions, or sometimes directly to two muons or pions \cite{Batell:2009yf, Essig:2009nc}.
Other hadronic final states are also possible for $m_{h_d}\gsim 1 \GeV$.
From \cite{Essig:2009nc, Batell:2009yf}, the production cross-sections are,
\bea
\sigma_{e^+e^-\rightarrow A'\gamma} &\approx& 60 \fb \ \parf{\epsilon^2c_W^2}{10^{-5}} \parf{E_{cm}^2-m_{A'}^2}{E_{cm}^2}^{-1},
\\
\sigma_{e^+e^-\rightarrow A'h_d} &\approx& 2 \fb \times \parf{\alpha_d}{\alpha} \parf{\epsilon^2c_W^2}{10^{-5}}
 \parf{E_{cm}^2-(m_{A'}+m_{h_d})^2}{E_{cm}^2}^{-1},
\eea
valid over the range $1 \GeV \lesssim m_{A'}+m_{h_d} < E_{cm}$.
See \cite{Essig:2009nc, Batell:2009yf} for a detailed discussion of these reactions. 

\begin{figure*}[!]
\begin{center}     
\includegraphics[width=.485\textwidth]{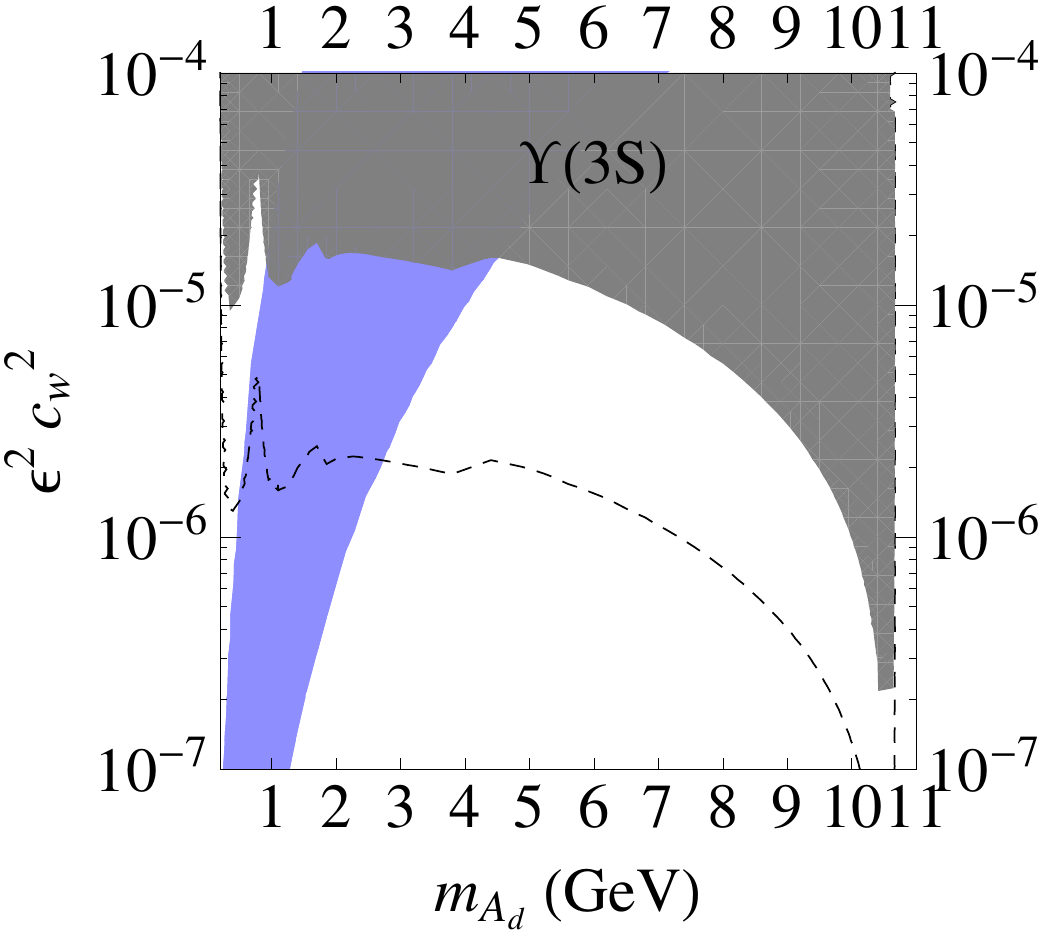}
\caption{Constraints derived from the $e^+e^-\rightarrow A' \gamma$ process in the $\mu^+\mu^-\gamma$ search 
in BaBar's $\Upsilon(3S)$ data using \cite{Aubert:2009cp}. The gray region is excluded 
at more than $90\%$ confidence, while the dotted line shows the combined reach of BaBar and Belle.  
The blue region shows the expected range of parameters in models of light sub-dominant higgsino dark matter.
While not shown, we expect that multi-lepton searches sensitive to the reaction 
$e^+e^-\rightarrow A' h_d$ in the four or more lepton/pion final state can reach sensitivities 
of order $(\epsilon c_W)^2 \sim \OO(10^{-8})$ \cite{:2009pw}.}
\label{fig:Bfactory}
\end{center}
\end{figure*}

Assuming that the branching fraction for the decay 
$A'\rightarrow \mu^+\mu^-$ is $1/(1+R_{\mu^+\mu^-})\lesssim 30\%$
for $m_{A'}\gsim 1 \GeV$, current limits on 
$e^+e^-\rightarrow A'\gamma\rightarrow \mu^+\mu^-\gamma$
require $(\epsilon c_W)^2 \lesssim 2\times 10^{-5}$ over the range 
$1 \GeV \lesssim m_{A'} \lesssim 6 \GeV$ \cite{Aubert:2009cp},
and are somewhat stronger at higher mass. 
Figure \ref{fig:Bfactory} illustrates the limit on $(\epsilon c_W)^2$ that we derived
from \cite{Aubert:2009cp}.
Limits on $e^+e^-\rightarrow 4l$ \cite{:2009pw}
and $e^+e^-\rightarrow \mbox{invisible}+\gamma$ \cite{:2009qd, :2010jw} can also be constraining,
but are slightly weaker than the di-lepton channel for the range of parameters we are considering. 
Likewise, rare decay constraints are also not as strong as the di-lepton search \cite{Batell:2009jf, Freytsis:2009ct}.

In the models considered in this paper, $(\epsilon c_W)^2$
is predicted as a function of $\alpha_D$ and $m_{A'}$.  
An improved search for $e^+e^-\rightarrow A' \gamma$ in the $\mu^+\mu^-\gamma$
final state using the full $\Upsilon(4S)$ data set at BaBar should be sufficient to probe almost 
the entire parameter space of this class of theories. 
Even better is a combined analysis of BaBar and Belle data in this channel. 
If kinematically accessible, the $e^+e^-\rightarrow A' h_d$ process can offer 
a particularly striking signature in the six lepton (and pion) final state. 
Existing searches in four lepton channels \cite{ :2009pw} indicate that upcoming searches in higher lepton
(and pion) multiplicity channels should thoroughly explore this region.

For the scenario where the relic abundance of higgsino GeV-Matter
arises from thermal freeze-out, normalizing $\alpha_D\propto
m_{A'}^2/m_{\tilde h}$ to the CoGeNT signal implies a particularly
sharp prediction for $(\epsilon c_W)^2$, given by
\eqref{ecw:annihilate}.  The predicted $B$-factory cross-sections in
this case are remarkably close to current limits \cite{Aubert:2009cp,
  :2009pw, :2009qd, :2010jw} and are readily tested by extending
existing $B$-factory searches.

While we believe that $B$-factories are uniquely suited to explore the light dark matter
scenarios outlined in this paper, LEP and hadronic collider searches for 
lepton jets can also play an important role.
If Standard Model super-partners are light enough to be directly produced at colliders, then 
supersymmetric dark 
sectors can give rise to spectacular lepton jet signals at the Tevatron or LHC
\cite{Strassler:2006im,Baumgart:2009tn, Cheung:2009su, Abazov:2009hn,Falkowski:2010cm}.
However, note that the super-partner scale can be as high as $\sim m_W/\sqrt{\epsilon}\sim 10 \TeV$ before irreducible super-symmetry breaking feeds into the dark sectors we consider. 
Additional searches in $J/\Psi$-factories \cite{Yin:2009mc,Li:2009wz}, $\phi$-factories \cite{Bossi:2009uw, AmelinoCamelia:2010me},
fixed-target experiments \cite{Bjorken:2009mm, Reece:2009un, Batell:2009di, Essig:2010xa, Freytsis:2009bh, Heinemeyer:2007sq, Schuster:2009au, Abouzaid:2006kk},
and satellite searches \cite{Batell:2009zp, Schuster:2009fc, Meade:2009mu} 
can provide evidence for (or constrain) GeV-scale gauge sectors coupled through kinetic mixing. 

\section{Conclusion}

If supersymmetry stabilizes the weak scale, then 
kinetic mixing with a new gauge force can naturally generate a GeV-scale
sector. This remarkable possibility could open the door
to exploring supersymmetry at low energies. 
In simple supersymmetric dark sectors, 
stable GeV-Matter can arise in the dark higgs sector,
and these may comprise a portion of the dark matter
with new and interesting phenomenology. 
This possibility is well-motivated by cosmic ray data suggestive of
GeV-scale forces.

In this paper, we have proposed simple models of GeV-Matter within a supersymmetric dark sector 
that couples to the Standard Model through kinetic mixing.
The low-energy effective Lagrangian we have studied has already appeared in the literature 
to realize inelastic up-scattering of $\OO(100 \GeV)$ mass dark matter as an explanation of the DAMA signal. 
Our interpretation of the data is different: 
\begin{itemize}
\item The CoGeNT and DAMA direct detection anomalies are simultaneously 
explained by inelastic {\it down-scattering} of $\OO(\GeV)$ mass dark-sector states (GeV-Matter), 
for which the higgsinos of a supersymmetric dark sector are an excellent candidate.
These states comprise a very sub-dominant fraction of the dark matter.  
Tension with existing experiments is reduced, and the CoGeNT and DAMA rates readily agree 
--- in contrast to elastic scattering explanations. 
\item Electroweak symmetry breaking triggers breaking of the dark sector $U(1)_D$, giving 
the vector multiplet a mass $m_{A'}\sim \sqrt{\epsilon g_D}m_W\sim \GeV$. The  direct detection cross-section of GeV-Matter is predicted (equation~\eqref{eq:DtermScatter}) in terms of Standard Model parameters with no dependence on $\epsilon$ or  $\alpha_D$.

\item The correct CoGeNT and DAMA event rates are obtained for GeV-Matter comprising 0.2--1\% of the halo mass density.  This abundance can be explained by thermal freeze-out, and the couplings required for this scenario are readily tested with existing $B$-factory data.
\item Another possibility is that a metastable WIMP decays to GeV-Matter after thermal GeV-Matter annihilation has frozen out. In this case, 
$\frac{\rho_{\rm light}}{\rho_{\rm WIMP}}\approx \frac{m_{\rm light}}{m_{\rm WIMP}}$ robustly produces the desired relic abundance for metastable WIMP masses of 2--3 TeV.  We have exhibited a model where the metastable WIMP is the bosonic superpartner of the TeV-mass dark matter whose decays can explain the cosmic ray data.
\end{itemize}

Thus, in our unified and predictive framework, the GeV-scale dark
sector plays a crucial role in explaining cosmic ray excesses {\it
  and} contains stable GeV-Matter responsible for direct detection signals.
The most important test of our interpretation of DAMA and CoGeNT will
come from further searches with direct detection experiments. In
particular, further studies using
CDMS Silicon and Germanium data, especially with lower threshold energy, will be vital.  
The power of these experiments depends sensitively on their threshold energies
and signal efficiencies.  Future studies should include these
uncertainties when quoting constraints.  Low-energy flavor factories
play an equally important role in testing our proposal.  A
resonance search in $e^+e^-\rightarrow \gamma \mu^+\mu^-$ in the
full BaBar or Belle datasets and higher lepton-multiplicity searches should be powerful enough to discover or exclude the light vector $A'$ over most of the parameter space that can
explain CoGeNT and DAMA.  Additional searches in
$J/\Psi$-factories \cite{Yin:2009mc}, $\phi$-factories
\cite{Bossi:2009uw, AmelinoCamelia:2010me}, and fixed-target
experiments \cite{Bjorken:2009mm, Reece:2009un, Batell:2009di,
  Essig:2010xa, Freytsis:2009bh, Heinemeyer:2007sq, Schuster:2009au}
can provide evidence for (or constrain) a GeV-scale gauge sector
coupled through kinetic mixing.  Finally, if Standard Model
super-partners are light enough to be directly produced at colliders,
then supersymmetric dark sectors can give rise to spectacular lepton
jet signals at the Tevatron or LHC \cite{ArkaniHamed:2008qp,Baumgart:2009tn,
  Cheung:2009su, Abazov:2009hn}.

\vspace{5mm}
\noindent
{\bf Acknowledgements}
\vspace{3mm}

We thank Clifford Cheung, Liam Fitzpatrick, Michael Peskin, and Jay Wacker for many useful discussions.  NT thanks Peter Graham, Roni Harnik, Surjeet Rajendran, and Prashant
Saraswat for early discussions of inelastic down-scattering at DAMA.  
We thank Spencer Chang, Jia Liu, Aaron Pierce, Neal Weiner, and Itay Yavin for bringing to our attention their related work 
\cite{Chang:2010yk}.  We also thank N. Weiner for alerting us to a quenching factor error in the first version of this paper, and for providing us with \cite{cogentNote}.  RE, JK,
and PS are supported by the US DOE under contract number
DE-AC02-76SF00515.


\appendix
\section{A UV completion for GeV-Matter from Decay}
\label{sec:uvcompletion}
For completeness, we present a simple UV completion of the model of Section \ref{sec:decay}, where the GeV-Matter states are populated from the decay of a TeV scale state.  
Consider the following matter content:
\begin{center}
\begin{tabular}{|c||c|c|c|c|c|c|c|c|c|c|c|}
\hline
                              & \ $S$ \ & \ $\Phi_u$ \ & \ $\Phi_d$ \ & \ $h_u$ \ & \ $h_d$ \ & \ $Z_1$ \ & \ $Z_2$ \ & \ $X$ \ & \ $Y$ \ & \ $A$ \ & \ $\bar{A}$ \ \\
\hline
\hline
$U(1)_d$             & 0 & 1 & -1 & 1 & -1 & 0 & 0 & 0 & 0 & -1 & 1 \\
\hline
$U(1)_R$             & 1 & 0 & 2 & 0 & 1 & 1 & 1 & $\varphi$ & $\varphi$ & $2-\varphi$ & $\varphi$ \\
\hline
$\mathbb{Z}_2$ & 0 & -1 & -1 & 0 & 0 & 0 & -1 & -1 & 0 & 0 & 0 \\
\hline
\end{tabular}
\end{center}
where $\varphi$ is a real number chosen in such a way to guarantee the form of the 
superpotential and K\"ahler potential corrections.   
At some very high scale, the complete superpotential is given by 
\be
{\cal W} = W_{\rm MSSM} + W_{\rm dark} + W_{\rm mix} + W_{\rm split} + W_{\rm K}  
+ W_{\rm break} , 
\ee 
where $W_{\rm MSSM}$ is the supersymmetric Standard Model,
\be
W_{\rm dark} = -\f{1}{4} W_D^2 + \lambda S h_u h_d + M \Phi_u \Phi_d
\ee
contains the GeV-scale dark sector fields (including the higgsinos, which will form the GeV-Matter) 
and the TeV-scale dominant dark matter component from the $\Phi_{u,d}$, and 
\be
W_{\rm mix} =  - \frac{\epsilon}{2} W^{\alpha}_Y W_{\alpha D}
\ee
contains the gauge-kinetic mixing term that connects the supersymmetric Standard Model to the dark sector.  
In addition, 
\be
W_{\rm split} =  Z_1 h_u h_d + Z_2 \Phi_u h_d - \f{1}{2} M_1 Z_1^2 - \f{1}{2} M_2 Z_2^2
\ee
will, after integrating out the $\sim$TeV-mass fields $Z_1$ and $Z_2$, generate the 
higher dimensional terms in 
Eq.~(\ref{eq:lowenergy}) that generate the mass splittings among the heavy and light dark matter states. 
The superpotential 
\be
W_{\rm K} =  A X \Phi_u + A Y h_u + M_* A \bar{A}
\ee
will, after integrating out the fields $A$ and $\bar{A}$ at some high scale $M_*$, 
generate the K\"ahler potential term Eq.~(\ref{eq:Kahler}), but no additional superpotential terms.  
Note that the symmetries guarantee that the K\"ahler potential has no lower-dimensional terms, 
or other terms of the same dimension as Eq.~(\ref{eq:Kahler}).  
Finally, we will not specify the precise form of $W_{\rm break}$.  This superpotential 
contains fields that break supersymmetry to give non-zero F-terms $F_X$ and $F_Y$.  
It also has to contain the supersymmetry breaking sector that generates the soft masses in the 
supersymmetric Standard Model via some mediation mechanism.   


\section{Sources of WIMP and GeV-Matter Mass Splittings}
\label{app:splittings}
We briefly remark on three models that give rise to inelastic
splittings for the $\Phi$ fermions and dark higgsinos.  The simplest model
has $\Phi_{u,d}$ charges $\pm 1/2$, so that additional interactions
\be
{\cal W'} = \lambda_+ h_d \Phi_u^2 + \lambda_- h_u \Phi_d^2 
\ee
are possible.  The former directly generates a somewhat large Majorana
mass for the heavy dark matter, while integrating out $\Phi$ loops or
other heavy states generates an operator
of the form \eqref{eqn:splitop} and therefore a splitting of the dark
higgsinos.

Smaller hierarchies of splittings are achieved in models where  $\Phi_{u,d}$ have charge $\pm 1$.  We first consider a model with two singlets $X$ and $Y$, and  two ${\mathbb Z}_2$ global symmetries: one under which $X$, $S$, and $h_u$ are odd and one under which $Y$, $S$, and $h_d$ are odd.  These symmetries allow, in addition to \eqref{eqn:lightW} and \eqref{eqn:heavyW}, relevant and marginal interactions 
\be
{\cal W}_{XY} = \mu_X X^2 +
\mu_Y Y^2 + \lambda_- \Phi_d h_u X + \lambda_+ \Phi_u h_d Y + \lambda_s S X Y.
\ee
Integrating out $X$ and $Y$ gives rise to interactions
\be
\frac{1}{M} \Phi_u^2h_d^2 \ \  \mbox{ and } \ \ \frac{1}{16 \pi^2 } \frac{1}{M} h_u^2h_d^2,
\ee
which split the heavy dark matter and dark higgsinos respectively.  The $\Phi$ splitting is generated at tree-level and the dark higgsino splitting only at loop-level, so the dark higgsino splitting is naturally smaller.  For Yukawa couplings $\sim 1/10$, $\langle h_d \rangle \sim 10$ GeV, and $M \sim 1$ TeV the typical splittings are 1 MeV and 10 keV respectively for the heavy and light states. 

Alternately, our $\Phi$ and $U(1)_D$ higgs-sector fields can couple to a doubly-charged field $T_{\pm 2}$ through interactions
\be
{\cal W}_{T}  = \mu_T T_{+2} T_{-2} + \lambda_d h_d^2 T_{+2} + \lambda_u h_u^2 T_{-2} + 
              \lambda_- \Phi_d^2 T_{+2} + \lambda_+ \Phi_u^2 T_{-2}. 
\ee
After $U(1)_D$ symmetry-breaking, the $F$-flat condition for $T_{+2}$ is satisfied if $T_{-2}$ acquires a small vev $\langle T_{-2} \rangle = - \lambda_d v_d^2/\mu_T \sim$ MeV for $\mu_T \sim$ few TeV.  This vev induces small Majorana masses for both $\Phi_d$ and $h_d$; a hierarchy of Yukawa couplings $\lambda_u \ll \lambda_+$ leads to a hierarchy of splittings.

\section{Form Factor and Halo Parametrizations for Direct Detection} 
\label{app:DirectDetection}
We elaborate here on the form factor and halo parametrizations
appearing in \eqref{eqn:dRdEr}, which are used throughout Section \ref{sec:DirectDetection}.  
$F(E_R)$ is a nuclear form factor, which we model as a Helm form factor as in \cite{Lewin:1995rx},
\be
F(q r_n) = \f{3 j_1(q r_n)}{q r_n} e^{-(q s)^2/2},
\ee
with $q = \sqrt{2 m_N E_R}$, $r_n^2 = c^2 + \f{7}{3}\pi^2 a^2 - 5 s^2$, 
$c = 1.23 A^{1/3} - 0.60~{\rm fm}$, $s=0.9$ fm, and $a=0.52$ fm. 

In the final factor of \eqref{eqn:dRdEr}, 
\be
 \int _{v_{\rm min}}^{v_{\rm esc}} d^3 v \f{f(\vec{v}-\vec{v}_e)}{v},
\ee
$\vec{v}$ is the dark matter velocity in the frame of the Earth, 
$\vec{v}_e$ is the velocity of the Earth in galactic coordinates, and 
$f(\vec{v}-\vec{v}_e)$ is the local dark matter velocity distribution in the Earth rest frame, 
which we model by 
\be
f(\vec{v}-\vec{v}_e) \propto \Big(e^{-(\vec{v}-\vec{v}_e)^2/v_0^2} - e^{-v_{\rm esc}^2/v_0^2}\Big), 
\ee
for $v<v_{\rm esc}$ and 0 otherwise.  
We normalize $f(\vec{v}-\vec{v}_e)$ to unity (for example, if $v_{\rm esc} = \infty$, then 
the proportionality constant is $1/(\pi v_0^2)^{3/2}$).
We use 
$v_0 = 254$ km/s \cite{Reid:2009nr} for both the average velocity of local stars and the mean velocity of halo particles.  The local galactic escape velocity $v_{esc}$ is measured to lie in the range 498 km/s $< v_{esc} <$ 608 km/s at 
90\% confidence \cite{Smith:2006ym}, and we use the median likelihood value of  $v_{esc} \simeq 544$ km/s.  The Earth velocity in the galactic frame is determined as a function of time by $\vec{v}_e = \vec{v}_\odot + \vec{v}_\oplus$, where $\vec{v}_\odot$ is the Sun's 
velocity relative to the Galactic rest frame \cite{Reid:2009nr,Dehnen:1997cq}, 
\be
\vec{v}_\odot = 
\left(\begin{array}{c} 10.00 \\ 5.23 \\ 7.17 \end{array} \right) \, {\rm km/s} +
\left(\begin{array}{c} 0 \\v_0 \\0 \end{array} \right) \,
\ee
and $\vec{v}_\oplus$ is the Earth's velocity relative to the Sun, 
\be
\vec{v}_\oplus = \langle u_E\rangle (1-e\sin(\lambda-\lambda_0))
\!
\left(\begin{array}{c} \cos(\beta_x) \sin(\lambda-\lambda_x) \\ 
			\cos(\beta_y) \sin(\lambda-\lambda_y) \\
			\cos(\beta_z) \sin(\lambda-\lambda_z) \end{array} \right) {\rm km/s}.
\ee
Here, the Earth's orbit has a mean velocity $\langle u_E\rangle = 29.79$ km/s and ellipticity 
$e=0.016722$, $\lambda$ is the angular position of the Earth's orbit, 
and the quantities $\beta_{x,y,z}$ and $\lambda_{0,x,y,z}$ are given in 
\cite{Lewin:1995rx}.  

The velocity integral in \eqref{eqn:dRdEr} cuts off at a minimum
velocity $v_{min}$ determined by the scattering kinematics:
\be
v_{\rm min} = \f{1}{\sqrt{2 m_N E_R}}\Big|\f{m_N E_R}{\mu_N} + \delta \Big|,
\ee 
where $\mu_N$ is the dark matter-nucleus reduced mass, and $\delta$ is
the splitting between the incoming and outgoing dark matter particles:
$\delta=0$ for elastic scattering, $\delta > 0$ if the incoming dark
matter particle scatters into a heavier excited state
(``up-scattering'') \cite{TuckerSmith:2001hy}, while $\delta < 0$ if
the incoming dark matter particle is already in an excited state and
scatters into a lighter ground state (``down-scattering'').  For the
masses and splittings we consider, at most a tiny fraction of the halo
can up-scatter and the down-scattering signal dominates.

\bibliography{Species}
\end{document}